\documentclass{aa}
\usepackage{graphicx, epsfig,psfig}
\usepackage{graphicx,subfigure,rotating}

\usepackage{natbib}
\bibpunct{(}{)}{;}{a}{}{,} 

\begin{document}

   \title{Analysis and modeling of high temporal resolution spectroscopic observations of flares on \mbox{\object{AD Leo}}}

   \titlerunning{Analysis and modeling of flares on AD~Leo}

   \author{I. Crespo-Chac\'on\inst{1}
          \and D. Montes\inst{1}
      \and D. Garc\'{\i}a-Alvarez\inst{2,3}
          \and M. J. Fern\'andez-Figueroa\inst{1}
          \and J. L\'opez-Santiago\inst{1,4}
          \and B. H. Foing\inst{5}
      }

   \offprints{icc@astrax.fis.ucm.es}

   \institute{Departamento de Astrof\'{\i}sica, Facultad de Ciencias
              F\'{\i}sicas, Universidad Complutense de Madrid, E-28040 Madrid, Spain\\
              \email{icc@astrax.fis.ucm.es}
         \and Harvard-Smithsonian Center for Astrophysics, 60 Garden Street, Cambridge, MA 02138, USA
         \and Armagh Observatory, College Hill, Armagh BT61 9DG, N. Ireland
         \and Osservatorio Astronomico di Palermo, Piazza del Parlamento 1, I-90134 Palermo, Italy
         \and Research Division, ESA Space Science Department, ESTEC/SCI-R, P.O. Box 299,
              2200 AG Noordwijk, The Netherlands
              }

   \date{Received ....., 2005; accepted ...., 2006}

   \abstract{We report the results of a high temporal resolution
spectroscopic monitoring of the flare star \mbox{\object{AD Leo}}. 
During 4 nights,
more than 600
spectra were taken in the optical range using the
Isaac Newton Telescope (INT) and the
Intermediate Dispersion Spectrograph (IDS). 
We have 
observed a large number of short and weak flares 
occurring very frequently (flare activity $>$ 0.71 hours$^{-1}$).
This is in 
favour of the very important role that flares can play in stellar coronal heating. 
The detected flares are non \mbox{white-light} flares and, though most of
solar flares belong to this kind, very few such events had been previously observed 
on stars.
The behaviour of different chromospheric lines 
(Balmer series from H$\alpha$ to H$_{11}$, \ion{Ca}{ii}~H \& K, 
\ion{Na}{i}~D$_1$ \& D$_2$, \ion{He}{i}~4026~\AA~and \ion{He}{i}~D$_3$) 
has been studied in detail for a total of 14 flares. 
We have also estimated the physical parameters of the flaring plasma 
by using a procedure which assumes a simplified slab
model of flares. 
All the obtained physical parameters are consistent with previously 
derived values for stellar flares, and the 
\mbox{areas -- less} than 2.3~$\%$ of the stellar \mbox{surface --} are comparable with the size
inferred for other solar and stellar flares. 
Finally, we have studied the relationships between the physical parameters
and the area, duration, maximum flux and energy released
during the detected flares.

   }

   \maketitle

\section{Introduction}

Stellar flares are events where a large amount of energy is released in a
short interval of time, taking place changes at almost all frequencies in
the electromagnetic spectrum. Flares are believed to be the result of the 
release of part of the magnetic energy stored in the corona through
 magnetic reconnection
\citep[see reviews by][]{Mirzoyan84,Haisch91,Garcia-Alvarez00}. 
However, the exact mechanisms leading to the energy
release and subsequent excitation of various emission features remain
poorly understood. Many types
of cool stars produce flares, sometimes at levels
several orders of magnitude more energetic than their solar counterparts
\citep{Pettersen89,Garcia-Alvarez02}. In dMe stars (UV Ceti-type stars) optical flares
are a common phenomenon. On the contrary, in more luminous stars flares are usually only
detected through UV or X-ray observations \citep{Doyle89b}, though
some optical flares have been observed in young early K dwarfs like LQ Hya
and PW And \citep{Montes99,jls03}.


\begin{table*}
\caption[]{Observing log INT/IDS (2 -- 5 April 2001). \label{tab:obslog}}
\begin{flushleft}
\scriptsize
\begin{tabular}{ccccccccccccccc}
\noalign{\smallskip} \hline  \hline \noalign{\smallskip}

&  \multicolumn{8}{c}{R1200B} & &  \multicolumn{5}{c}{R1200Y} \\

\noalign{\smallskip} \cline{2-9} \cline{11-15} \noalign{\smallskip}

Night & $N$ & UT & $t_{\rm exp}$(s)
 & \multicolumn{5}{c}{$SNR$} & & $N$ & UT
 & $t_{\rm exp}$(s) & \multicolumn{2}{c}{$SNR$}\\

 & & start--end & min--max & \multicolumn{5}{c}{min--max} &
 & & start--end & min--max & \multicolumn{2}{c}{min--max}\\

\noalign{\smallskip} \cline{5-9} \cline{14-15} \noalign{\smallskip}

 & & & & H$\beta$ & H$\gamma$ & H$\delta$ & \ion{Ca}{ii} H & H$_8$ &
 & & & & H$\alpha$ &  \ion{Na}{i} D$_{1}$ \\

 & & & & & & & \ion{Ca}{ii} K & H$_9$ & & & & & &  \ion{Na}{i} D$_{2}$ \\

 & & & & & & & & H$_{10}$ & & & & & &  \ion{He}{i} D$_{3}$\\

\noalign{\smallskip} \cline{1-15} \noalign{\smallskip}

1 & 18 & 00:02--01:47 & 180--300 & 76--85 & 53--58 & 45--49
  & 33--37 & 28--31 & & 0 & -- & -- & -- & -- \\

2 & 91 & 20:19--01:58 & 60--120 & 46--73 &  32--48 & 26--40
  & 20--29 & 17--25 & & 0 & -- & -- & -- & --\\

3 & 246  & 20:10--02:37 & 15--120 & 33--89 &  21--58 & 18--51
  & 13--38 & 12--31 & & 0 & -- & -- & -- & --\\

4 & 104 & 20:28--22:44 & 15--120 & 27--72 & 18--48 & 15--41
  & 11--30 & 10--25 & & 189 & 23:38--03:13 & 5--120
  &  32--172 & 23--114\\

\noalign{\smallskip}
\cline{1-15} \noalign{\smallskip}
\end{tabular}
\end{flushleft}

\end{table*}

One would like to trace all the energetic processes in a flare to
a common origin, though the released energy can be very different.
The largest solar flares
involve energies of $10^{32}\ \mathrm{ergs}$
\citep{Gershberg89}. Large flares on dMe stars can be two
orders of magnitude larger \citep{Doyle90a,Byrne90}, while
very energetic flares are produced by \mbox{RS CVn} binary systems,
where the total
energy may exceed
$10^{38}\ \mathrm{ergs}$
\citep{Doyle92,Foing94,Garcia-Alvarez02b}. Such a change in the
star's radiation field modifies drastically the atmospheric
properties over large areas, from photospheric to coronal layers.
{Models by \citet{Houdebine92} indicate that heating may be propagated
down to low photospheric levels, with densities higher than
$10^{16}\ \mathrm{cm}^{-3}$. However, electrons with energies in
the $\mathrm{MeV}$ range would be required to attain such depth.

Spectral emission lines are the most appropriate diagnostics to constrain the
physical properties and motions of flaring plasmas 
\citep[see][and references therein]{Houdebine03}. Unfortunately, only a few sets
of observations with adequate time and spectral resolution are so
far available \citep{Rodono89,Hawley&Pettersen91,Houdebine92,Gunn1994,Garcia-Alvarez02,Hawley03}.
Furthermore, there are also few attempts to derive the physical parameters of the flare plasmas  
\citep{Donati-Falchi85,Kunkel70,Gershberg74,Katsova90,Jevremovic98}.
\citet{Garcia-Alvarez02} obtained the first detailed trace of physical parameters during a
large optical flare, on \mbox{\object{AT Mic}}, using the \citet{Jevremovic98} procedure.
Good quality spectroscopic observations, as those analyzed in this work for \mbox{\object{AD Leo}},
and a correct description of the different flare
components are required to constrain numerical simulations.

\mbox{\object{AD Leo}} is well-known for being a frequent source of flares. It has
been subject of numerous studies
in the optical, EUV and X-rays because of its nature as one of
the most active M dwarfs \citep[][]{Pettersen84,Pettersen90,Lang86,Gudel89,Hawley&Pettersen91,Hawley95,Hawley03,Abada-Simon97,Cully97,Favata00,jsf2002,vandenBesselaar2003,Maggio04,Robrade05,Smith05}.
\mbox{\object{AD Leo}} (\mbox{\object{GJ 388}}) is classified as dM3Ve \citep{Henry94}.
This star has an unseen \mbox{companion -- detected} 
using speckle \mbox{interferometry -- with} 
a period of about \mbox{27~years} which
is expected to have a very low mass \citep{Balega84}.
\mbox{\object{AD Leo}} is
located in the immediate solar neighbourhood, at a distance of \mbox{$\sim$~4.9~pc}
(from the ground-based parallax \mbox{\citep{Gray91} -- it} was not a
Hipparcos target). Its high activity is probably due to its high
rotation rate \citep[\mbox{$P_{\rm phot}$~$\sim$~$2.7$~days},][]{Spiesman86}. In fact, its
rotational velocity (\mbox{$v {\rm sin}i$ = 6.2~$\pm$~0.8~km~s$^{-1}$})
places \mbox{\object{AD Leo}} among the tail of rare fast rotating \mbox{M dwarfs} \citep{Delfosse98}.
The radius and mass of \mbox{\object{AD Leo}} are
\mbox{$R$~$\sim$~0.44$R_{\odot}$} and \mbox{$M$~$\sim$~0.40$M_{\odot}$}, respectively
\citep{Pettersen76,Favata00}. At this
mass, stars are expected to have a substantial
radiative core, so that the interior structure is still ``solar-like'' \citep{Chabrier97}.
\citet{Saar85} detected, through infrared line measurements,
photospheric fields on \mbox{\object{AD Leo}} showing the presence of strong magnetic
fields.
They inferred that 73~\% of the \mbox{\object{AD Leo's}} surface is covered by active
regions outside of dark spots containing a mean field strength of \mbox{$B$ =
3800~$\pm$~260~G}.

Here we present the results of a long spectroscopic monitoring of \mbox{\object{AD Leo}}
that was carried out using an intermediate dispersion spectrograph and
high temporal resolution. This work has
considerably extended the existing sample of stellar flares analyzed with good quality
spectroscopy in the optical range.
Details about the technical information of the observations and 
data reduction are given in $\S$~\ref{sec:dataset}.
Section~\ref{sec:analysis} describes the analysis of the
observations and the detected flares, including equivalent widths, line fluxes,
released energy, line profiles and asymmetries.
In $\S$~\ref{sec:flaresmodelling} we present the main physical plasma
parameters obtained for the observed flares using the code developed by
\citet{Jevremovic98}. The discussion of the results and
conclusions are given in~$\S$~\ref{sec:conclusions}, where the relationships
between the physical and observational parameters are also analyzed.
The preliminary results for this star and \mbox{\object{V1054 Oph}} were
presented by \citet{Montes03} and \citet{icc04}.

\begin{figure*}
\begin{center}
{\psfig{figure=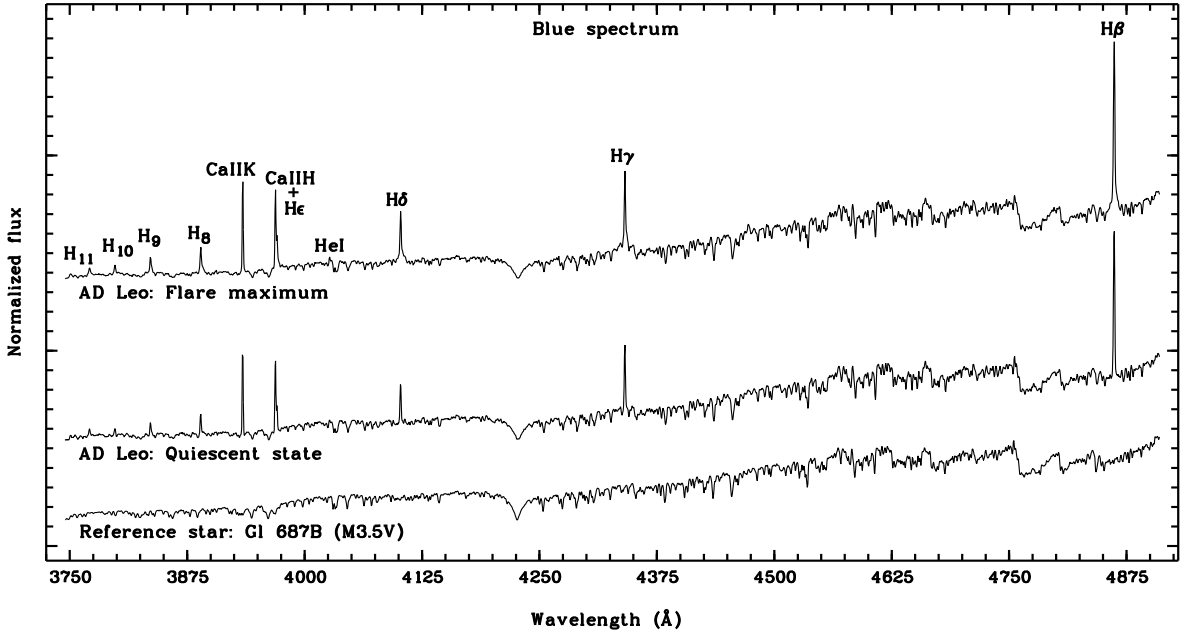,bbllx=34pt,bblly=67pt,bburx=536pt,bbury=343pt,height=8.4cm,width=15.8cm,clip=}}
\caption[ ]{Observed spectrum of \mbox{\object{AD Leo}} at the maximum of the strongest flare detected with the R1200B grating (flare 2, see $\S$~\ref{sec:equivalentwidths}) and in its
quiescent state. The observed spectrum of the reference star \mbox{\object{Gl 687B}} 
is shown at the bottom.
The chromospheric lines are identified.
\label{fig:espectros_Hb} }
\end{center}
\end{figure*}

\begin{figure*}
\begin{center}
{\psfig{figure=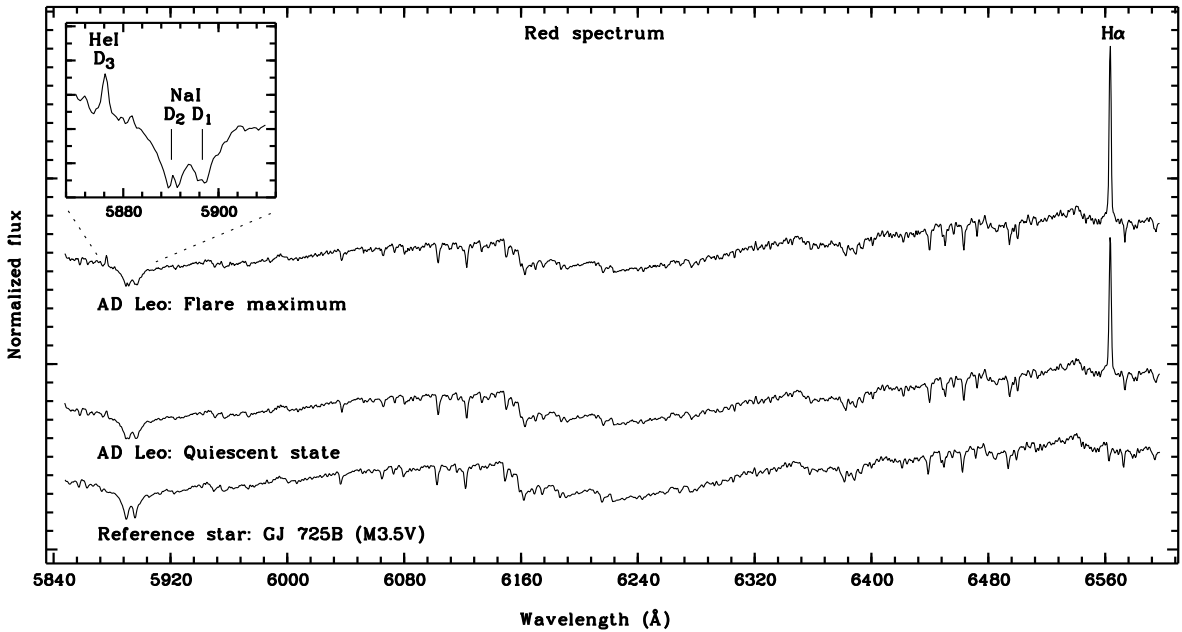,bbllx=34pt,bblly=67pt,bburx=536pt,bbury=343pt,height=8.4cm,width=15.8cm,clip=}}
\caption[ ]{As Fig.~\ref{fig:espectros_Hb} but for the 
spectra taken with the R1200Y grating, the strongest flare detected using this spectral configuration (flare 12, see $\S$~\ref{sec:equivalentwidths}), and the
reference star \mbox{\object{GJ 725B}}.
\label{fig:espectros_Ha} }
\end{center}
\end{figure*}

\section{Observations and data reduction}
\label{sec:dataset}

The data 
were taken during the MUlti-SIte COntinuous Spectroscopy 
(MUSICOS) 2001 campaign.
It involved observations at two sites:
El Roque de los Muchachos Observatory from La Palma (Spain) 
and SAAO (South Africa).
However, due to poor weather conditions, the only useful
data were those taken at La Palma.
This observing run was carried out
with the 2.5~m Isaac Newton Telescope (INT)
during \mbox{2 -- 5} April 2001.
The Intermediate Dispersion Spectrograph (IDS) was utilized together with the
2148x4200 EEV10a CCD detector.
Two gratings were used: R1200B (every night) and
R1200Y (second half of the last night).
The wavelength covered by R1200B (blue spectrum) ranges from
3554~\AA \ to 5176~\AA \ (including the Balmer lines from H$\beta$ to H$_{11}$
as well as the \ion{Ca}{ii}~H~\&~K and \ion{He}{i} 4026~\AA \ lines).
The reciprocal dispersion of these spectra is 0.48~\AA/pixel.
The wavelength covered by R1200Y (red spectrum) ranges from
5527~\AA \ to 7137~\AA \ (including the H$\alpha$,
\mbox{\ion{Na}{i} D$_{1}$} \& D$_{2}$ and \mbox{\ion{He}{i} D$_{3}$ lines}),
being the reciprocal dispersion 0.47~\AA/pixel. The spectral
resolution, determined as the full width at half maximum (FWHM) of
the arc comparison lines, is 1.22~\AA~for the blue region and
1.13~\AA~for the red one.

\begin{figure*}[t!]
\begin{center}
{\psfig{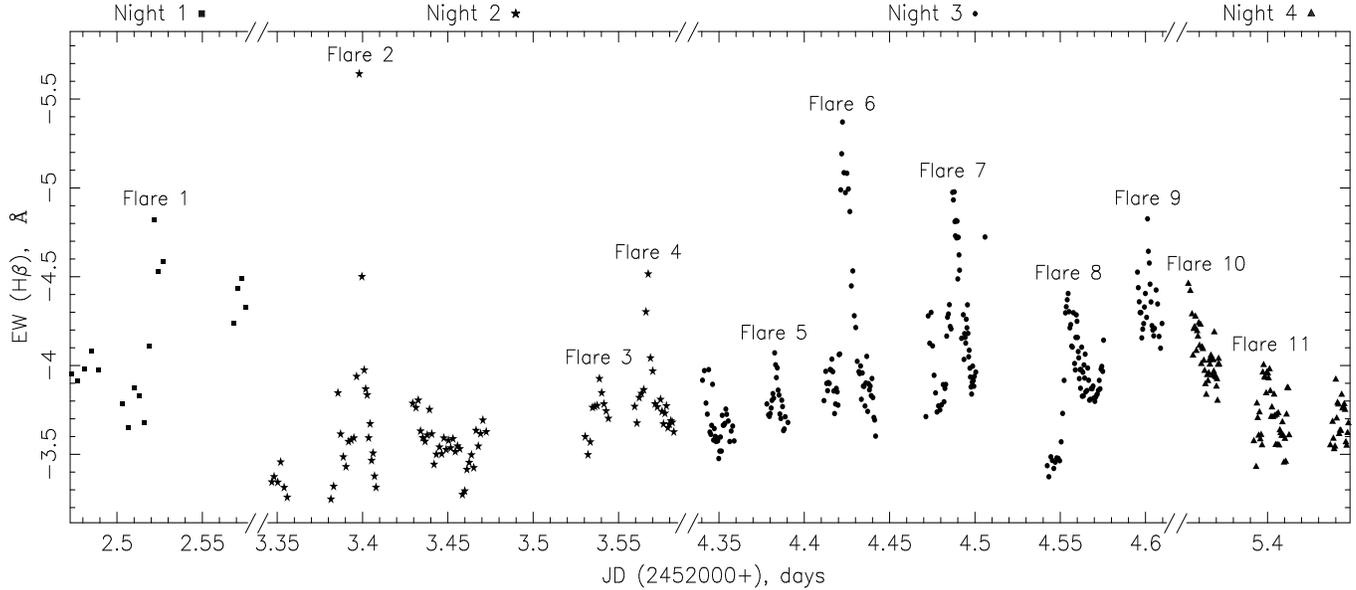}}
\caption[ ]{Temporal evolution of the $EW$ of the H$\beta$ line.
The strongest observed flares are labeled.
\label{fig:adleo_JD_EW_4noches} }
\end{center}
\end{figure*}

\begin{figure}
\begin{center}
{\psfig{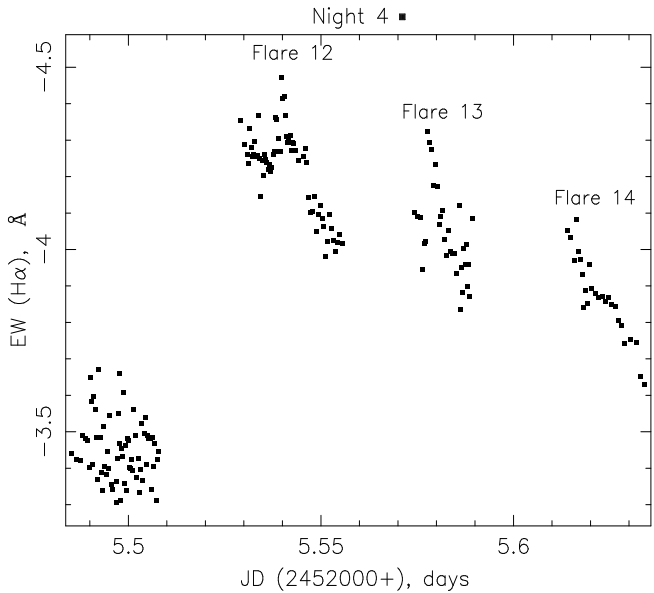}}
\caption[ ]{As Fig.~\ref{fig:adleo_JD_EW_4noches} but for the H$\alpha$ line.
The observed flares have been numbered following the order in Fig.~\ref{fig:adleo_JD_EW_4noches}.
\label{fig:adleo_JD_EW_Ha} }
\end{center}
\end{figure}

We took series of spectra with short exposure times:
from 15 to 300 s for R1200B and from 5 to 120 s
for R1200Y. The spectra in each series were separated 
only by the CCD readout time (less than 60 s)
in order to obtain the highest temporal resolution as possible.
During 4 nights, a total of 459 spectra of \mbox{\object{AD Leo}} were
obtained with R1200B and 189 with R1200Y. 
The observing log (Table~\ref{tab:obslog}) lists: the number of spectra
taken every night ($N$), the universal time (UT)
at the beginning and end of the \mbox{\object{AD Leo}} observations, the minimum and
maximum exposure time ($t_{\rm exp}$) of the spectra, and the \mbox{signal-to-noise ratio} ($SNR$) of
the continuum near each line region.
These quantities are given for both gratings (R1200B and R1200Y).
Note that, despite the short exposure times, the $SNR$
is large enough to perform a reliable analysis.

The reduction has been done following the standard procedure:
bias and dark subtraction,
flat-field correction using exposures of a tungsten lamp, cosmic rays correction,
and sky background subtraction from the region of the aperture
chosen for doing an optimal extraction of each spectrum.
The software packages of IRAF\footnote{IRAF is distributed by the National
    Optical Astronomy Observatories,
    which are operated by the Association of Universities for Research
    in Astronomy, Inc., under cooperative agreement with the National
    Science Foundation.} have been used.
The wavelength calibration has been done by using spectra of Cu--Ar lamps.
Finally, all the spectra have been normalized to their maximum flux value of the
observed continuum.

\section{Analysis of the observations}
\label{sec:analysis}

In Fig.~\ref{fig:espectros_Hb} we have plotted the observed blue spectrum of \mbox{\object{AD Leo}} in its
quiescent state (minimum observed emission level) and at the maximum of the strongest
flare 
detected 
with the R1200B grating (flare 2, see $\S$~\ref{sec:equivalentwidths}).
Fig.~\ref{fig:espectros_Ha} is similar to Fig.~\ref{fig:espectros_Hb} but
for the red spectra, taken with the R1200Y grating, and the strongest flare
detected with this spectral configuration (flare 12, see $\S$~\ref{sec:equivalentwidths}).
In addition, the spectra of \mbox{\object{Gl 687B}} and \mbox{\object{GJ 725B}} have
been plotted as examples of inactive stars 
with a spectral type and luminosity class (M3.5V) very similar to those of 
\mbox{\object{AD Leo}} (M3V). Note that the origin of the Y-axis is different for every
spectrum in order to avoid the overlapping between them.


\begin{table*}
\begin{center}
\caption[]{Minimum and maximum relative error in the $EW$ of the different chromospheric lines.
\label{tab:relative_error}}
\scriptsize
\begin{tabular}{ccccccccccccccc}
\noalign{\smallskip}
\hline  \hline
\noalign{\smallskip}

Night & \multicolumn{14}{c}{$\bigtriangleup EW_{\rm rel}$ (\%)}\\
 & \multicolumn{14}{c}{(min--max)}\\
\noalign{\smallskip}
\cline{2-15}
\noalign{\smallskip}
 & H$\beta$ & H$\gamma$ & H$\delta$ & \ion{He}{i} $\lambda$4026\AA & \ion{Ca}{ii} K &
\ion{Ca}{ii} H+H$\epsilon$ & H$_{8}$ & H$_{9}$ & H$_{10}$ & H$_{11}$ &
H$\alpha$ &
\ion{Na}{i} D$_{1}$ & \ion{Na}{i} D$_{2}$ & \ion{He}{i} D$_{3}$\\
\noalign{\smallskip}
\cline{1-15}
\noalign{\smallskip}

1 & 10--13 & 6--9  & 8--14 & 20--46 & 5--8 & 5--8 &
  9--20 & 9--17 & 11--24 & 16--37 & -- & -- & -- & -- \\
2 &  9--15 & 6--10 & 7--17 & 19--57 & 5--9 & 5--10 &
  9--23 & 9--22 & 12--34 & 17--62 & -- & -- & -- & -- \\
3 & 10--17 & 6--12 & 8--20 & 24--109 & 5--13 & 5--13 &
  10--28 & 9--29 & 12--41 & 17--61 & -- & -- & -- & -- \\
4 & 10--20 & 6--15 & 8--23 & 27--152 & 5--14 & 6--15 &
  10--34 & 10--39 & 12--56 & 18--145 & 9--20 & 2--5 & 1--5 & 32--328 \\

\noalign{\smallskip}
\cline{1-15}
\noalign{\smallskip}
\end{tabular}
\end{center}

\end{table*}

\mbox{\object{AD Leo}} shows a high level of chromospheric activity.
A strong emission in the Balmer series and the
\ion{Ca}{ii}~H~\&~K lines is observed even in its quiescent state
(see Fig.~\ref{fig:espectros_Hb} and~\ref{fig:espectros_Ha}). The
greatest emission in these lines is found at the maximum of the detected flares.
The \mbox{\ion{He}{i} D$_3$} and  \mbox{\ion{He}{i} 4026~\AA}~lines also show emission above the
continuum, which is more noticeable during the strongest flares.
The \ion{Na}{i} D$_1$~\&~D$_2$ absorption lines only present a
slight filled-in profile in the quiescent state, though
a low emission is observed in their core during flares.
High resolution spectroscopic observations of the quiescent state of \mbox{\object{AD Leo}} confirm
the presence of chromospheric emission lines that range from a
weak emission in the center of the \ion{Ca}{ii} IRT and the
\ion{Na}{i} D$_{1}$ \& D$_{2}$ lines to a strong emission in
the \ion{He}{i} D$_{3}$, \ion{Mg}{ii} h \& k, \ion{Ca}{ii}~H~\&~K and \ion{H}{i}
Balmer lines \citep{Pettersen81,Doyle87,icc05}. In addition, \citet{Sundland88}
suggested that Balmer lines are the most important components to 
chromospheric radiation loss.

\mbox{\object{AD Leo}} is well-known for having strong flares \citep[see, for example,][]{Hawley&Pettersen91}.
However, after overlapping the normalized spectra 
and comparing the depth of the absorption lines and the shape of the continuum,
we have not detected any noticeable continuum change (see also
Fig.~\ref{fig:espectros_Hb} and~\ref{fig:espectros_Ha}).
Therefore the flares analyzed in this work are non \mbox{white-light} flares
and, within the studied wavelength range, they only affect the emission in
the chromospheric lines. Despite this kind of flares is the most typical in
the Sun, in which no detectable signature in \mbox{white-light} is generally observed,
very few such events had been previously detected on stars
\citep[see][]{Butler86,Houdebine03}.
Conversely to the Sun, stellar non \mbox{white-light} flares have been seldom observed
because very little time has been dedicated to spectroscopy in comparison 
to photometry.
However, \citet{Houdebine92} suggested that the low frequency observed 
for stellar non \mbox{white-light} flares
could be due to a strong contrast effect. In other words,
many flares have been detected as \mbox{white-light} flares on dMe stars
because of the relative weakness of their photospheric background, but
they would have been classified as non \mbox{white-light} flares on the Sun because of
its much higher photospheric background.

\subsection{Equivalent widths}
\label{sec:equivalentwidths}

The equivalent width ($EW$) of the observed chromospheric lines has
been measured very carefully in order to detect possible weak flares in our observations.
Two methods have been tried out.
The first one consists of obtaining the $EW$ with the routine SPLOT included
in IRAF, taking exactly the same wavelength limits for each line in all the spectra.
The second method uses the routine SBANDS of IRAF,
which generally introduces less noise in the $EW$ measurement 
because of taking into account a larger number of points to calculate the continuum value. 
For this reason, we only present the results
obtained with SBANDS.
Note that the line regions have been taken
wide enough to include the enhancement of the line wings during
flares. 

SBANDS uses Eq.~\ref{eq:EW2} to obtain the $EW$ of a line; where
$W_{{\rm l}}$ is the width at the base of the line,
$d$ is the reciprocal dispersion (pixel size in~\AA),
${F_{{\rm c}, {\rm i}}}$ is the flux per pixel in the continuum under the line,
${F_{{\rm l}, {\rm i}}}$ is the observed flux in the pixel i of the line,
and $n$ is the number of pixels within the line region.
%

{\small
\begin{equation}
EW = {W_{{\rm l}} - \frac{d}{F_{{\rm c}, {\rm i}}} \ \displaystyle{\sum_{{\rm i}=1}^{{\rm i}=n} F_{{\rm l}, {\rm i}}}}
\label{eq:EW2}
\end{equation}}

The $EW$ uncertainty has been estimated using Eq.~\ref{eq:EW2error}. We have
considered \mbox{$\bigtriangleup W_{{\rm l}}=0~{\rm \AA}$} and \mbox{$\bigtriangleup d=0.01~{\rm \AA/pixel}$}.
Eq.~\ref{eq:EW2error}
has been obtained by applying the standard quadratic error propagation theory
to Eq.~\ref{eq:EW2}, taking into account that \mbox{$n=W_{{\rm l}}/d$} and
assuming
\mbox{$\bigtriangleup F_{{\rm l}, {\rm i}}\approx \bigtriangleup F_{{\rm c}, {\rm i}} = F_{{\rm c}, {\rm i}}/{\rm SNR}$}.

{\small
\begin{equation}
\bigtriangleup EW = \sqrt{\left(C_1\right)^2+
\left(C_2\right)^2+
\left(C_3\right)^2+
\left(C_4\right)^2}
\label{eq:EW2error}
\end{equation}}

where
 
{\small
\[C_1 = \frac{\partial EW}{\partial W_{{\rm l}}}\bigtriangleup W_{{\rm l}}
= \bigtriangleup W_{{\rm l}}\]}
 
{\small
\[C_2 = \frac{\partial EW}{\partial d}\bigtriangleup d
= \frac{EW-W_{{\rm l}}}{d}\bigtriangleup d\]}
 
{\small
\[C_3 = \frac{\partial EW}{\partial F_{{\rm c}, {\rm i}}}\bigtriangleup F_{{\rm c}, {\rm i}}
= \frac{W_{{\rm l}} - EW}{{\rm SNR}}\]}
 
{\small
\[C_4 = \frac{\partial EW}{\partial F_{{\rm l}, {\rm i}}}\bigtriangleup F_{{\rm l}, {\rm i}}
= \frac{\sqrt{W_{{\rm l}} \ d}}{{\rm SNR}}\]}


\begin{table*}
\begin{center}
\caption[]{$JD$ of the flare onset ($JD_{\rm start}$), total duration
of the detected flares and length of their phases
(seen in the H$\beta$ line in the case of 
flares 1 to 11 and
seen in the H$\alpha$ line in the case of
flares 12 to 14). The time delay
between the maximum of the 
chromospheric lines and the one of 
H$\beta$ 
or H$\alpha$, depending on the spectral configuration,
is also shown. Blanks are given when
the beginning, maximum and/or end of the flare was not observed in the
line under consideration.
\label{tab:flares1to11_duration}}
\scriptsize
\begin{tabular}{ccccccccccccccc}
\noalign{\smallskip}
\hline  \hline
\noalign{\smallskip}

Flare & $JD_{\rm start}$ (days) & \multicolumn{3}{c}{Duration (min)} & & \multicolumn{2}{c}{Delay of the
maximum (min)} \\
& (2452000+) & & & & & & \\
\noalign{\smallskip}
\cline{3-5}
\cline{7-8}
\noalign{\smallskip}
 & & Total & Impulsive phase & Gradual decay & & ~~\ion{Ca}{ii} K & \ion{He}{i}~4026~\AA \\
 & & (H$\beta$) & (H$\beta$) & (H$\beta$) & & & \\
\noalign{\smallskip}
\cline{1-8}
\noalign{\smallskip}

1 & 2.513$\pm$0.003 & -- & 13$\pm$7 & -- & & ~~-- & 0$\pm$6 \\
2 & 3.3903$\pm$0.0016 & 25$\pm$4 & 11$\pm$5 & 14$\pm$4 & & ~~0$\pm$5 & 0$\pm$5 \\
3 & 3.5321$\pm$0.0016 & -- & 10$\pm$4 & -- & & ~~0$\pm$4 & 0$\pm$4 \\
4 & 3.5608$\pm$0.0013 & 31$\pm$3 & 10$\pm$4 & 22$\pm$3 & & ~~0$\pm$4 & 0$\pm$4 \\
5 & 4.3781$\pm$0.0010 & 18$\pm$4 & 6$\pm$2 & 11$\pm$3 & & ~~0$\pm$1 & -- \\
6 & 4.4193$\pm$0.0005 & 22$\pm$1 & 4$\pm$2 & 17$\pm$2 & & ~~5$\pm$3 & 0$\pm$3 \\
7 & 4.4804$\pm$0.0010 & 25$\pm$2 & 9$\pm$3 & 16$\pm$2 & & ~~3$\pm$2 & 1$\pm$2 \\
8 & 4.5479$\pm$0.0009 & -- & 10$\pm$2 & -- & & ~~1$\pm$1 & 1$\pm$1 \\
9 & 4.5978$\pm$0.0005 & 17$\pm$2 & 5$\pm$1 & 12$\pm$2 & & ~~2$\pm$1 & -- \\
10 & -- & -- & -- & -- & & ~~-- & -- \\
11 & 5.3968$\pm$0.0005 & 14$\pm$1 & 1$\pm$1 & 12$\pm$1 & & ~~3$\pm$1 & -- \\

\noalign{\smallskip}
\cline{1-8}
\noalign{\smallskip}
\end{tabular}
\begin{tabular}{ccccccccccccccc}

Flare & $JD_{\rm start}$ (days) & \multicolumn{3}{c}{Duration (min)} & & \multicolumn{2}{c}{Delay of the
maximum (min)} \\
& (2452000+) & & & & & & \\
\noalign{\smallskip}
\cline{3-5}
\cline{7-8}
\noalign{\smallskip}
 & & Total & Impulsive phase & Gradual decay & & ~~\ion{Na}{i} D$_1$, D$_2$  & \ion{He}{i} D$_{3}$ \\
 & & (H$\alpha$) & (H$\alpha$) & (H$\alpha$) & & & \\
\noalign{\smallskip}
\cline{1-8}
\noalign{\smallskip}
12 & 5.5341$\pm$0.0003 & 31$\pm$2 & 5.9$\pm$0.9 & 25$\pm$2 & & ~~0.9$\pm$0.9 & -- \\
13 & 5.5763$\pm$0.0005 & 14$\pm$1 & 2$\pm$1 & 12$\pm$1 & & ~~1$\pm$1 & 1$\pm$1 \\
14 & -- & -- & -- & -- & & ~~-- & -- \\

\noalign{\smallskip}
\cline{1-8}
\noalign{\smallskip}
\end{tabular}
\end{center}
\end{table*}

Table~\ref{tab:relative_error} shows the minimum and maximum values
of the relative error in $EW$ ($\bigtriangleup EW_{\rm rel}$)
for each chromospheric line and every night.
The $\bigtriangleup EW_{\rm rel}$ of the \ion{Ca}{ii}~H~\&~K, \ion{Na}{i}~D$_{1}$ \& D$_{2}$,
H$\beta$, H$\gamma$ and H$\delta$ lines is between
1~\% and 20~\%. For H$_{\rm 8}$ and H$_{\rm 9}$ the 
$\bigtriangleup EW_{\rm rel}$ is less than 40~\%, while
for H$_{\rm 10}$ and H$_{\rm 11}$
is higher than the uncertainty estimated for the other Balmer lines
(up to 50~\% in the case of H$_{\rm 10}$ 
and even larger than 100~\% in the case of H$_{\rm 11}$).
In addition, the $\bigtriangleup EW_{\rm rel}$ of \mbox{\ion{He}{i} D$_3$} and 
\mbox{\ion{He}{i} 4026~\AA}~is frequently greater than
100~\%. For this reason, the results obtained
for the 
\ion{He}{i} lines are less reliable.
Besides, because of the short exposure times and the 
relative faintness of \mbox{\object{AD Leo}} in the blue,
the region \mbox{blueward} H$_{\rm 10}$ is usually too noisy for reliable measurements.

The observed flares have been detected using
the H$\beta$ and H$\alpha$ lines (for the blue and red spectra,
respectively). H$\beta$ and H$\alpha$
have been chosen because they suffer from large variations during flares
and are the chromospheric lines with the best $SNR$ in each
spectral configuration.

Fig.~\ref{fig:adleo_JD_EW_4noches} shows
the temporal evolution found for the $EW$ of the H$\beta$ line.
The observed flares are marked with numbers.
Other smaller changes can also be seen.
The temporal evolution found for the $EW$ of the H$\alpha$
line is given in Fig.~\ref{fig:adleo_JD_EW_Ha}.
No strong variations have been observed
during our spectroscopic monitoring. However,
14 short and weak flares have been detected: 11 among the blue spectra and
3 among the red ones. The Julian Date ($JD$) at the beginning of each
flare ($JD_{\rm start}$) is shown in Table~\ref{tab:flares1to11_duration}.
These flares have been also observed in the other chromospheric lines.
Nevertheless, it is more difficult to distinguish between flares 
and noise for lower $SNR$ and/or weaker lines.
In addition,
a modulation with a period of $\sim$~2~days 
is also observed in the quiescent emission of \mbox{\object{AD Leo}}, though the
photometric period of this star was found to be 2.7~days \citep{Spiesman86}.

\begin{table*}
\caption[]{$EWRQ$ at flare maximum ($EWRQ_{\rm max}$) for the different chromospheric lines.
Blanks are given when the flare maximum was not detected in the line under consideration.
Flares with no maximum data in any line have been omitted.
\label{tab:EWrelativevariation}}
\scriptsize
\begin{tabular}{ccccccccccccccc}
\noalign{\smallskip}
\hline  \hline
\noalign{\smallskip}

Flare & \multicolumn{10}{c}{$EWRQ_{\rm max}$} \\
\noalign{\smallskip}
\cline{2-11}
\noalign{\smallskip}
 & H$\beta$ & H$\gamma$ & H$\delta$ & H$_{8}$ & H$_{9}$ &
   H$_{10}$ & H$_{11}$ & \ion{Ca}{ii} H + H$\epsilon$ &
   \ion{Ca}{ii} K & \ion{He}{i}~4026~\AA \\
\noalign{\smallskip}
\cline{1-11}
\noalign{\smallskip}

1 & 1.26$\pm$0.18 & 1.28$\pm$0.11 & 1.47$\pm$0.17 & 1.53$\pm$0.22 & 1.52$\pm$0.20 & 1.7$\pm$0.3 & 1.6$\pm$0.4 & 1.16$\pm$0.09 & -- & 1.5$\pm$0.5 \\

2 & 1.65$\pm$0.25 & 1.52$\pm$0.14 & 2.0$\pm$0.3 & 2.0$\pm$0.3 & 1.8$\pm$0.3 & 1.5$\pm$0.3 & 2.0$\pm$0.7 & 1.30$\pm$0.11 & 1.14$\pm$0.09 & 1.7$\pm$0.6 \\

3 & 1.09$\pm$0.19 & 1.06$\pm$0.10 & 1.10$\pm$0.17 & 1.11$\pm$0.21 & 1.13$\pm$0.22 & 0.99$\pm$0.23 & 1.0$\pm$0.4 & 1.1$\pm$0.1 & 1.01$\pm$0.09 & 1.3$\pm$0.5 \\

4 & 1.23$\pm$0.20 & 1.25$\pm$0.12 & 1.44$\pm$0.21 & 1.5$\pm$0.3 & 1.5$\pm$0.3 & 1.6$\pm$0.3 & 1.6$\pm$0.6 & 1.13$\pm$0.10 & 1.06$\pm$0.09 & 1.2$\pm$0.5 \\

5 & 1.08$\pm$0.20 & 1.08$\pm$0.13 & 1.14$\pm$0.21 & 1.1$\pm$0.3 & 1.1$\pm$0.3 & 1.0$\pm$0.3 & 1.5$\pm$0.6 & 1.06$\pm$ 0.13 & 1.01$\pm$0.12 & -- \\

6 & 1.4$\pm$0.3 & 1.34$\pm$0.17 & 1.6$\pm$0.3 & 1.8$\pm$0.5 & 1.4$\pm$0.4 & 1.5$\pm$0.5 & 1.3$\pm$0.5 & 1.23$\pm$0.16 &1.14$\pm$0.13 & 1.6$\pm$0.9 \\

7 & 1.32$\pm$0.24 & 1.32$\pm$0.17 & 1.5$\pm$0.3 & 1.4$\pm$0.4 & 1.7$\pm$0.4 & 1.5$\pm$0.5 & 1.9$\pm$0.8 & 1.15$\pm$0.15 & 1.09$\pm$0.14 & 1.9$\pm$1.1 \\

8 & 1.27$\pm$0.23 & 1.25$\pm$0.14 & 1.49$\pm$0.25 & 1.5$\pm$0.3 & 1.5$\pm$0.3 & 1.4$\pm$0.4 & 1.5$\pm$0.6 & 1.16$\pm$0.13 & 1.06$\pm$0.11 & 1.6$\pm$0.8 \\

9 & 1.16$\pm$0.21 & 1.07$\pm$0.13 & 1.18$\pm$0.21 & 1.2$\pm$0.3 & 1.3$\pm$0.3 & 1.1$\pm$0.3 & 1.0$\pm$0.4 & 1.13$\pm$0.14 & 1.06$\pm$0.14 & -- \\


11 & 1.12$\pm$0.22 & 1.10$\pm$0.15 & 1.19$\pm$0.24 & 1.3$\pm$0.3 & 1.1$\pm$0.3 & 1.2$\pm$0.4 & 1.0$\pm$0.5 & 1.05$\pm$0.13 & 1.07$\pm$0.13 & -- \\

\noalign{\smallskip}
\cline{1-11}
\noalign{\smallskip}
\end{tabular}

\begin{tabular}{ccccccccccccccc}
Flare & \multicolumn{4}{c}{$EWRQ_{\rm max}$} \\
\noalign{\smallskip}
\cline{2-5}
\noalign{\smallskip}
 & H$\alpha$ & \ion{Na}{i} D$_{1}$ & \ion{Na}{i} D$_{2}$ & \ion{He}{i} D$_{3}$ \\
\noalign{\smallskip}
\cline{1-5}
\noalign{\smallskip}

12 & 1.09$\pm$0.15 & 0.95$\pm$0.04 & 0.95$\pm$0.03 & -- \\

13 & 1.13$\pm$0.18 & 0.98$\pm$0.04 & 0.98$\pm$0.04 & 1.5$\pm$1.1 \\


\noalign{\smallskip}
\cline{1-5}
\noalign{\smallskip}
\end{tabular}

\end{table*}

We want to emphasize 
that the flare frequency is very high,
and sometimes new flares take place when others are
still active (see, for example, 
\mbox{flares 8} and 11 and the points before
\mbox{flares 2} and 7 in Fig.~\ref{fig:adleo_JD_EW_4noches}).
It is also remarkable that the flares in \mbox{night 4} seem to erupt
over the gradual decay phase of other stronger flare
(Fig.~\ref{fig:adleo_JD_EW_Ha}).
For calculating the flare frequency we have taken the total observed time
as the sum of \mbox{UT$_{\rm end}$ -- UT$_{\rm start}$} 
for all the nights (see Table~\ref{tab:obslog}). 
Note that, within a night, the series of observations 
are not continuous in time. Therefore the flare frequency
should be greater than the obtained one. This implies that the flare activity of
\mbox{\object{AD Leo}}, found from the variations of the chromospheric emission lines,
\mbox{is $>$~0.71~flares/hour}. 
This value is slightly higher than those
that other authors measured by using photometry:
\citet{Moffett74} observed 0.42~flares/hour, 
\citet{Pettersen84} detected 0.57~flares/hour,
and \citet{Konstantinova1995} found \mbox{0.33 - 0.70~flares/hour}.
Given that the detected flares are non \mbox{white-light} flares,
we conclude that non \mbox{white-light} flares may be more frequent on stars
than \mbox{white-light} flares, as observed in the Sun.
In fact, as the total time of the gaps
between the series of observations is similar to the total real observed time,
the flare activity of \mbox{\object{AD Leo}} could even be the double of that
obtained. As far as we know, this is the first time that such a high flare frequency
is inferred from the variation of the chromospheric emission lines. 


\subsection{Equivalent widths relative to the quiescent state}
\label{sec:EWRQ}

In order to compare the behaviour of the different chromospheric lines,
we have used the equivalent width relative
to the quiescent state ($EWRQ$), defined as the
ratio of the $EW$ to the $EW$ in the quiescent state.
Fig.~\ref{fig:EWRQflares_a} and~\ref{fig:EWRQflares_b} show
the $EWRQ$ of several lines for a representative sample of the detected flares
(those in which all the phases have been observed). The chromospheric lines plotted in
these two figures are those with less uncertainty in the $EW$ measurement.
We can observe two different kinds of flares:
for some of them (see 
\mbox{flares 6},
9, 11, 12 and 13) the gradual decay phase of the Balmer
lines is much longer than the impulsive phase; in contrast,
other flares (2, 4, 5 and 7) are less impulsive.
This resembles the classification of solar flares
made by \citet{Pallavicini77}: {\it eruptive flares} or {\it long-decay events} and
{\it confined} or {\it compact flares}.
However, all these
flares always show the same behaviour
in the \mbox{\ion{Ca}{ii}~H \& K} lines, that is a slow evolution throughout all the event.
The evolution
of the \ion{Na}{i}~D$_{1}$ \& D$_{2}$, \mbox{\ion{He}{i} D$_3$} and 
\mbox{\ion{He}{i}} 4026~\AA~lines
is quite similar to that of the Balmer series. However, it is
less clear for the \ion{He}{i} lines because of having
a very high error sometimes.

\begin{figure*}
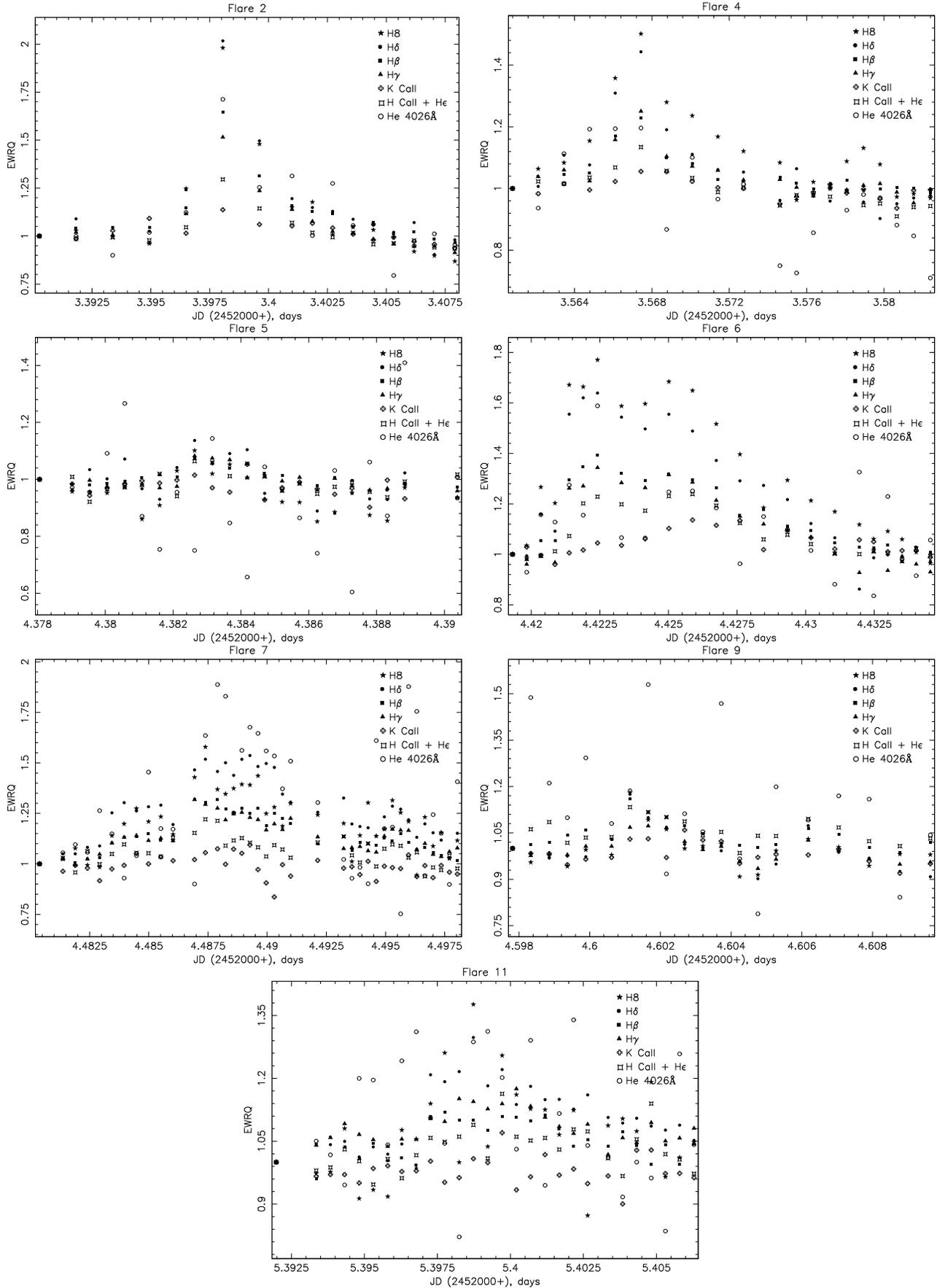

\begin{center}
{\psfig{figure=adleo_VRQEW_fulg2_publicacion1.ps,width=8.3cm}}~~
{\psfig{figure=adleo_VRQEW_fulg4_publicacion1.ps,width=8.3cm}}
{\psfig{figure=adleo_VRQEW_fulg5_publicacion1.ps,width=8.3cm}}~~
{\psfig{figure=adleo_VRQEW_fulg6_publicacion1.ps,width=8.3cm}}
{\psfig{figure=adleo_VRQEW_fulg7_publicacion1.ps,width=8.3cm}}~~
{\psfig{figure=adleo_VRQEW_fulg9_publicacion1.ps,width=8.3cm}}
{\psfig{figure=adleo_VRQEW_fulg11_publicacion1.ps,width=8.3cm}}
\caption[ ]{$EWRQ$ of the flares (detected using the R1200B grating)
in which all the phases have been observed 
(lines: H$\beta$, H$\gamma$, H$\delta$, H$_8$, \ion{Ca}{ii} H + H$\epsilon$,
\ion{Ca}{ii} K, \ion{He}{i} 4026~\AA).
\label{fig:EWRQflares_a}}
\end{center}
\end{figure*}

\begin{figure*}
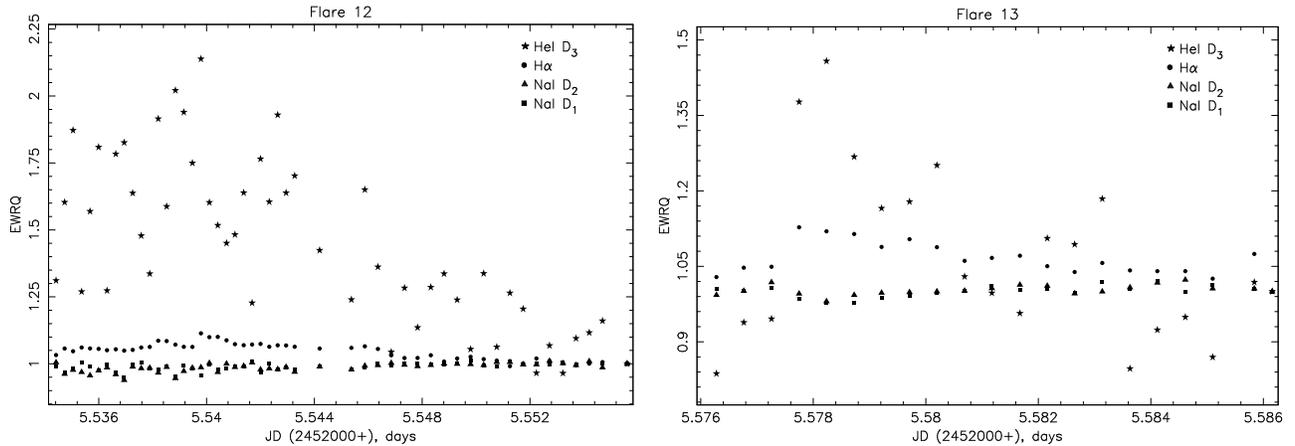

\begin{center}
{\psfig{figure=adleo_VRQEW_fulg12_publicacion1.ps,width=8.3cm}}~~
{\psfig{figure=adleo_VRQEW_fulg13_publicacion1.ps,width=8.3cm}}
\caption[ ]{As Fig.~\ref{fig:EWRQflares_a} but for the flares detected
using the R1200Y grating (lines: H$\alpha$, \mbox{\ion{Na}{i} D$_1$}, \mbox{\ion{Na}{i} D$_2$},
\mbox{\ion{He}{i} D$_3$}).
\label{fig:EWRQflares_b}}
\end{center}
\end{figure*}

In some flares (5, 6, 11, 12 and 13) we can see two maxima in the
\ion{Na}{i} and Balmer lines, but it is
not so evident for the \mbox{\ion{He}{i}}
and \mbox{\ion{Ca}{ii}} lines. Sometimes
variations at shorter time scales
are also detected during the gradual decay phase, in which different peaks,
decreasing in intensity,
are observed in the $EWRQ$ (see, for instance, 
flare 12). 
This could be interpreted as the
succession of different reconnection processes,
decreasing in efficiency, within the same flare
\mbox{-- following} the original suggestion of \mbox{\citet{Kopp76} --}
or as a series of flares
originated as a result of wave disturbances coming along the
stellar surface from the first flare region.

Table~\ref{tab:flares1to11_duration}
contains: the
total duration of the detected flares, the length of
their impulsive and gradual decay phases, and the time delay between the
maximum of each chromospheric line and that of H$\beta$ or H$\alpha$
(depending on the spectral configuration). The detected flares last from
14~$\pm$~1 up to 31~$\pm$~3~min.
Regarding 
the first maximum of emission, we have found that the Balmer series 
reach it simultaneously
while the rest of the lines are delayed.
The delay is negligible for the \mbox{\ion{He}{i}}
and \ion{Na}{i} lines
($\sim$~1~min) but it is very evident for
\ion{Ca}{ii}~H \&~K (up to 5~$\pm$~3~min).
It seems that this delay is greater for the flares with a shorter
impulsive phase of the Balmer lines.
However, the total duration of the flare does not seem to be related.
For the first five flares, we cannot assume 
that the delay of the maximum is completely zero
because their uncertainties, particularly those
of \ion{He}{i}~4026~\AA, are even greater than the delay 
found in the other cases.
The moment in which a line reaches its maximum is related to the
height where the line is formed above the stellar surface. This is
also related to the temperature that characterizes the formation
of the line. According to the line formation models in stellar atmospheres,
the \ion{Ca}{ii}~H~\&~K lines are formed at deeper and cooler layers
than the Balmer series. Therefore, the gas that is heated and evaporated
into the newly formed loop after magnetic reconnection \citep{Cargill83,Forbes86} cools and reaches
the formation temperature of the Balmer series before the one of
the \ion{Ca}{ii}~H~\&~K lines.
\citet{Houdebine03} found that the rise and decay times
in the \ion{Ca}{ii}~K and H$\gamma$ lines
obey good relationships, which
implies that there is a well-defined underlying
mechanism responsible for the flux time profiles in these lines.

Table~\ref{tab:EWrelativevariation} lists the $EWRQ$ of 
the chromospheric lines
at their maximum in each flare ($EWRQ_{\rm max}$). 
The lower the wavelength, the greater $EWRQ_{\rm max}$ is found for the Balmer lines.
However, H$\beta$ and H$\gamma$ have a very similar variation.
For \ion{He}{i}~4026~\AA~and \ion{He}{i}~D$_{3}$  
the $EWRQ_{\rm max}$ is analogous to that of the Balmer 
series, whereas for the \ion{Ca}{ii}~H \&~K and 
\ion{Na}{i}~D$_{1}$ \&~D$_{2}$ lines is smaller.
Our results show 
that the duration of flares tends to be larger when the
$EWRQ_{\rm max}$ of the Balmer lines is greater. This effect is
more noticeable for the less impulsive flares. 
Nevertheless, there are no clear relationships between
the $EWRQ_{\rm max}$ of the other lines and the duration of the flare.

\subsection{Line fluxes and released energy}
\label{sec:linefluxes}

In order to estimate the flare energy released in the observed
chromospheric lines, we have converted the $EW$ into absolute
surface fluxes and luminosities.

The absolute line fluxes ($F$) have been computed from the $EW$ measured
for each line and its local continuum. The absolute flux of the continuum
near each line has been determined {making use of} the method given
by~\citet{PettersenHawley89}. For \mbox{\object{AD Leo}},
they interpolate between the R and I filter photometric fluxes to
estimate the observed flux in the continuum near 8850~\AA.
This value is then transformed into a continuum surface flux of
\mbox{$F$({\rm 8850~\AA}) = 6.5 $\times$ 10$^{5}$~erg~s$^{-1}$cm$^{-2}$\AA$^{-1}$},
using 0.44~$R_{\odot}$ for the radius and 0.203~arcsec for the parallax.
Direct scaling from the flux-calibrated spectrum of \mbox{\object{AD Leo}},
showed by~\citet{PettersenHawley89},
has allowed us to estimate the continuum surface fluxes near the
emission lines of interest. We multiply this continuum value by the measured $EW$ to determine
the line surface flux. The absolute fluxes of the Balmer lines
at the different flare maxima ($F_{\rm max}$) are listed in
Table~\ref{tab:maximumfluxes}.
These values will be used in $\S$~\ref{sec:flaresmodelling}
to calculate the Balmer decrements.
Note that each $F_{\rm max}$ only shows the
contribution of the flare to the surface flux 
(the contribution of the quiescent state has been subtracted).
Since our aim is to
isolate the contribution of the flares from the total emission,
we have chosen the quiescent state for each of them as a pseudo-quiescent,
that is the minimum emission level just before or after the flare.
In this way, variations originated by other processes, which can contribute
to the observed emission, are not taken into account.
Even if these processes were magnetic reconnections
within the same active region, assuming that
the solar flare model is also applicable to the stellar case,
the new emitting
plasma would be placed in a different post-flare loop
\citep{Cargill83,Forbes86}.
Therefore, in first approximation, 
its physical properties (see $\S$~\ref{sec:flaresmodelling})
could be separately studied.

We have also converted the absolute fluxes into luminosities ($L$).
Table~\ref{tab:totalluminosity} lists the contribution of every flare 
to the energy released in the Balmer lines. The released energy
has been calculated by numerically
integrating the luminosity from the beginning to the end of the flare
and subtracting the contribution of its corresponding \mbox{pseudo-quiescent} state.
The energy released during the
impulsive and gradual decay phases, which has been calculated in the same way,
is given as well.


\begin{table*}
\begin{center}
\caption[]{Surface fluxes of the Balmer lines at flare maximum ($F_{\rm max}$).
The contribution of the quiescent state has been subtracted.
Flares with no maximum data in any line have been omitted.
\label{tab:maximumfluxes}}
\scriptsize
\begin{tabular}{ccccccccccccccc}
\noalign{\smallskip}
\hline  \hline
\noalign{\smallskip}

Flare & \multicolumn{8}{c}{$F_{\rm max}$ (10$^{5}$~erg~s$^{-1}$cm$^{-2}$)} \\
\noalign{\smallskip}
\cline{2-9}
\noalign{\smallskip}
 & H$\alpha$ & H$\beta$ & H$\gamma$ & H$\delta$ & H$_{8}$ & H$_{9}$ &
   H$_{10}$ & H$_{11}$ \\
\noalign{\smallskip}
\cline{1-9}
\noalign{\smallskip}

1  & -- & 0.90 & 0.64 & 0.56 & 0.29 & 0.23 & 0.109 & 0.070 \\
2  & -- & 1.82 & 1.10 & 0.90 & 0.47 & 0.33 & 0.108 & 0.105 \\
3  & -- & 0.34 & 0.19 & 0.15 & 0.08 & 0.053 & 0.011 & 0.013 \\
4  & -- & 0.70 & 0.52 & 0.41 & 0.22 & 0.17 & 0.120 & 0.079 \\
5  & -- & 0.34 & 0.23 & 0.19 & 0.10 & 0.04 & 0.079 & 0.061 \\
6  & -- & 1.19 & 0.70 & 0.62 & 0.32 & 0.16 & 0.084 & 0.040 \\
7  & -- & 0.94 & 0.62 & 0.43 & 0.19 & 0.25 & 0.076 & 0.090 \\
8  & -- & 0.74 & 0.48 & 0.42 & 0.16 & 0.14 & 0.072 & 0.053 \\
9  & -- & 0.57 & 0.27 & 0.30 & 0.11 & 0.11 & 0.070 & 0.036 \\
11 & -- & 0.35 & 0.07 & 0.18 & 0.08 & 0.03 & 0.050 & 0.010 \\
12 & 1.10 & -- & -- & -- & -- & -- & -- & -- \\
13 & 1.40 & -- & -- & -- & -- & -- & -- & -- \\

\noalign{\smallskip}
\cline{1-9}
\noalign{\smallskip}
\end{tabular}
\end{center}
\end{table*}



\begin{table*}
\begin{center}
\caption[]{Energy released in the Balmer lines during every flare phase. 
The total released energy is also shown. 
The contribution of the quiescent state has been subtracted.
We have omitted the flares with no data about their beginning, maximum and end.
\label{tab:totalluminosity}}
\scriptsize
\begin{tabular}{clccccccccccccc}
\noalign{\smallskip}
\hline  \hline
\noalign{\smallskip}

Flare & Phase & \multicolumn{8}{c}{Released energy (10$^{29}$~erg)} \\
\noalign{\smallskip}
\cline{3-10}
\noalign{\smallskip}
 & & H$\alpha$ & H$\beta$ & H$\gamma$ & H$\delta$ & H$_{8}$ & H$_{9}$ &
   H$_{10}$ & H$_{11}$ \\
\noalign{\smallskip}
\cline{1-10}
\noalign{\smallskip}

1 & Impulsive& -- & 2.5 & 1.2 & 1.6 & 0.8 & 0.7 & 0.3 & 0.2 \\
  & Gradual & -- & -- & -- & -- & -- & -- & -- & -- \\
  & Total & -- & -- & -- & -- & -- & -- & -- & -- \\

2 & Impulsive & -- & 3.3 & 2.3 & 1.4 & 0.9 & 0.9 & 0.2 & 0.3\\
  & Gradual & -- & 4.9 & 3.1 & 2.0 & 1.2 & 1.2 & 0.4 & 0.5 \\
  & Total & -- & 8.2 & 5.4 & 3.4 & 2.2 & 2.1 & 0.6 & 0.7 \\

3 & Impulsive & -- & 1.2 & 0.7 & 0.7 & 0.3 & 0.2 & 0.1 & 0.0 \\
  & Gradual & -- &  -- & -- & -- & -- & -- & -- & -- \\
  & Total & -- & -- & -- & -- & -- & -- & -- & -- \\

4 & Impulsive & -- & 1.9 & 1.3 & 0.9 & 0.6 & 0.5 & 0.4 & 0.3\\
  & Gradual & -- & 2.2 & 1.6 & 0.6 & 0.9 & 0.5 & 0.5 & 0.5 \\
  & Total & -- & 4.1 & 3.0 & 1.5 & 1.5 & 1.1 & 0.9 & 0.8 \\

5 & Impulsive & -- & 0.6 & 0.2 & 0.3 & 0.2 & 0.1 & 0.3 & 0.2 \\
  & Gradual & -- & 1.0 & 0.6 & 0.4 & 0.2 & 0.1 & 0.4 & 0.2 \\
  & Total & -- & 1.5 & 0.8 & 0.7 & 0.4 & 0.2 & 0.7 & 0.4 \\

6 & Impulsive & -- & 1.5 & 0.7 & 0.9 & 0.5 & 0.2 & 0.1 & 0.0 \\
  & Gradual & -- & 6.2 & 3.3 & 3.1 & 1.8 & 1.0 & 0.3 & 0.1 \\
  & Total & -- & 7.7 & 3.9 & 4.0 & 2.3 & 1.2 & 0.4 & 0.1 \\

7 & Impulsive & -- & 1.8 & 1.2 & 1.1 & 0.4 & 0.5 & 0.2 & 0.2 \\
  & Gradual & -- & 5.4 & 3.4 & 3.2 & 1.3 & 0.8 & 0.6 & 0.6 \\
  & Total & -- & 7.2 & 4.6 & 4.4 & 1.7 & 1.2 & 0.8 & 0.9 \\

8 & Impulsive & -- & 2.0 & 1.3 & 1.0 & 0.4 & 0.4 & 0.2 & 0.1 \\
  & Gradual & -- &  -- & -- & -- & -- & -- & -- & -- \\
  & Total & -- & -- & -- & -- & -- & -- & -- & -- \\

9 & Impulsive & -- & 0.6 & 0.4 & 0.3 & 0.0 & 0.0 & 0.1 & 0.0 \\
  & Gradual & -- & 1.5 & 0.9 & 0.7 & 0.1 & 0.1 & 0.3 & 0.2 \\
  & Total & -- & 2.1 & 1.3 & 1.0 & 0.1 & 0.1 & 0.5 & 0.2 \\


11 & Impulsive & -- & 0.2 & 0.1 & 0.1 & 0.0 & 0.0 & 0.0 & 0.0 \\
   & Gradual & -- & 1.8 & 0.6 & 1.2 & 0.2 & 0.2 & 0.4 & 0.1 \\
   & Total & -- & 2.0 & 0.7 & 1.3 & 0.2 & 0.2 & 0.5 & 0.1 \\

12 & Impulsive & 3.1 & -- & -- & -- & -- & -- & -- & --\\
   & Gradual & 10.3 & -- & -- & -- & -- & -- & -- & --\\
   & Total & 13.4 & -- & -- & -- & -- & -- & -- & -- \\

13 & Impulsive & 1.0 & -- & -- & -- & -- & -- & -- & -- \\
   & Gradual & 6.5 & -- & -- & -- & -- & -- & -- & -- \\
   & Total & 7.5 & -- & -- & -- & -- & -- & -- & -- \\


\noalign{\smallskip}
\cline{1-10}
\noalign{\smallskip}
\end{tabular}
\end{center}
\end{table*}

\subsection{Line profiles and asymmetries}

We focus this section on the Balmer series and 
\ion{Ca}{ii}~H \&~K lines. The intermediate spectral 
resolution of the observations has not allowed us
to study the profile of the \ion{He}{i} 
and \ion{Na}{i} lines because of their low intensity.

We have plotted the
temporal evolution of the observed line profiles during the detected flares
(see the examples given in Fig.~\ref{fig:profiles}).
No evidence for line-shifts has been found.
However, the reported flares may be too weak for 
measuring line-shifts using the available spectral resolution
\mbox{(see $\S$~\ref{sec:dataset})}.
It seems that flares affect the core of the lines
before the wings. The emission in both of them (core and wings)
increases during the impulsive phase,
shows the largest value when the $EWRQ$ of the line reaches the maximum,
and decreases during the gradual decay.
The emission in the wings
of the \mbox{\ion{Ca}{ii}~H \&~K} lines hardly rises during the observed
flares, being only noticeable during the strongest ones (see 
Fig.~\ref{fig:profiles}).
In fact, their broadening is negligible even at flare maximum.
If the flare is too weak, as the flare 5,
the profile of the Balmer lines shows the same
behaviour as that found for the \mbox{\ion{Ca}{ii}~H \&~K} lines.
In contrast, when the released energy is high enough
to produce changes in the wings of the Balmer series
(flares 2, 6 and 7),
the core of these lines seems to rise up to a constant value (which
is independent of the $EWRQ_{\rm max}$)
whereas the emission in the wings depends
on the flare energy.
Table~\ref{tab:widthatthebase} lists the width at the base of the
lines of interest during
the quiescent state and the maximum of the strongest
observed flare (\mbox{flare 2}).
The fact that
the width of the \ion{Ca}{ii} lines is much less sensitive to flares
than that of the Balmer lines suggests that the
broadening may be due to the Stark effect
\citep{Byrne1989,Robinson1989}.
Nevertheless, mass motions can also be present (see below).

\begin{figure}
\begin{center}
{\psfig{figure=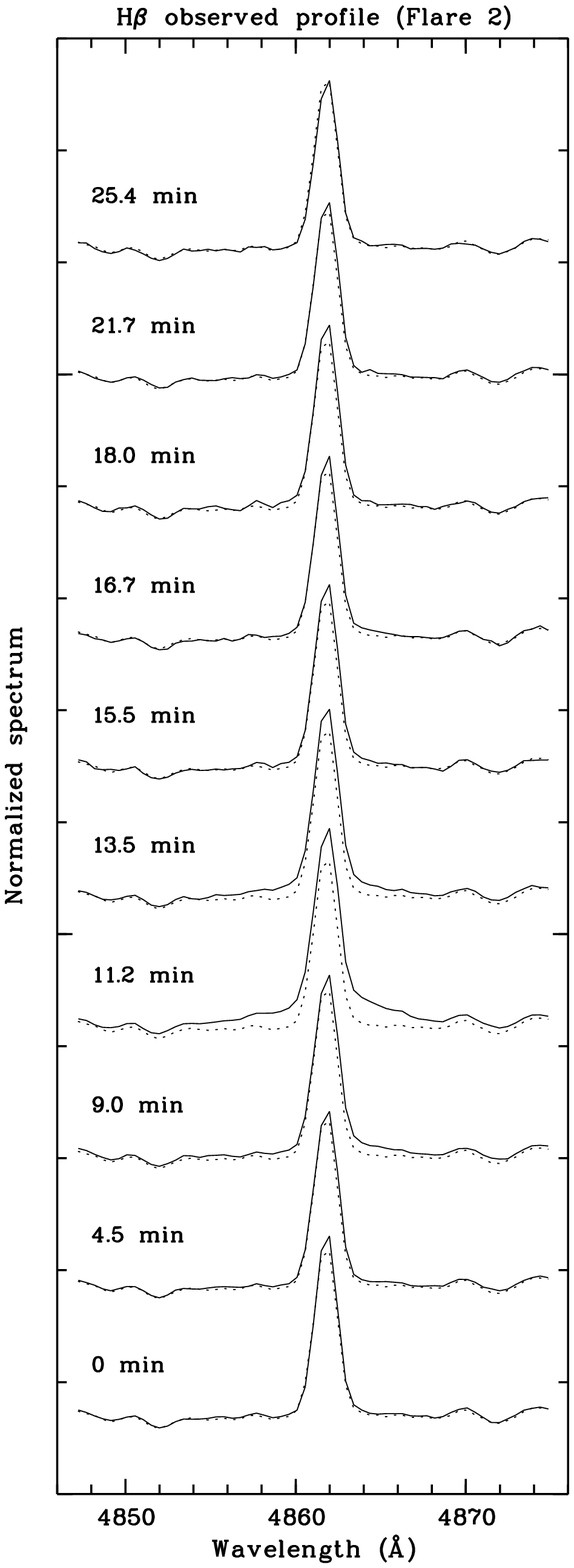,height=14.5cm,width=4.5cm,bbllx=28pt,bblly=28pt,bburx=303pt,bbury=773pt,clip=}}
{\psfig{figure=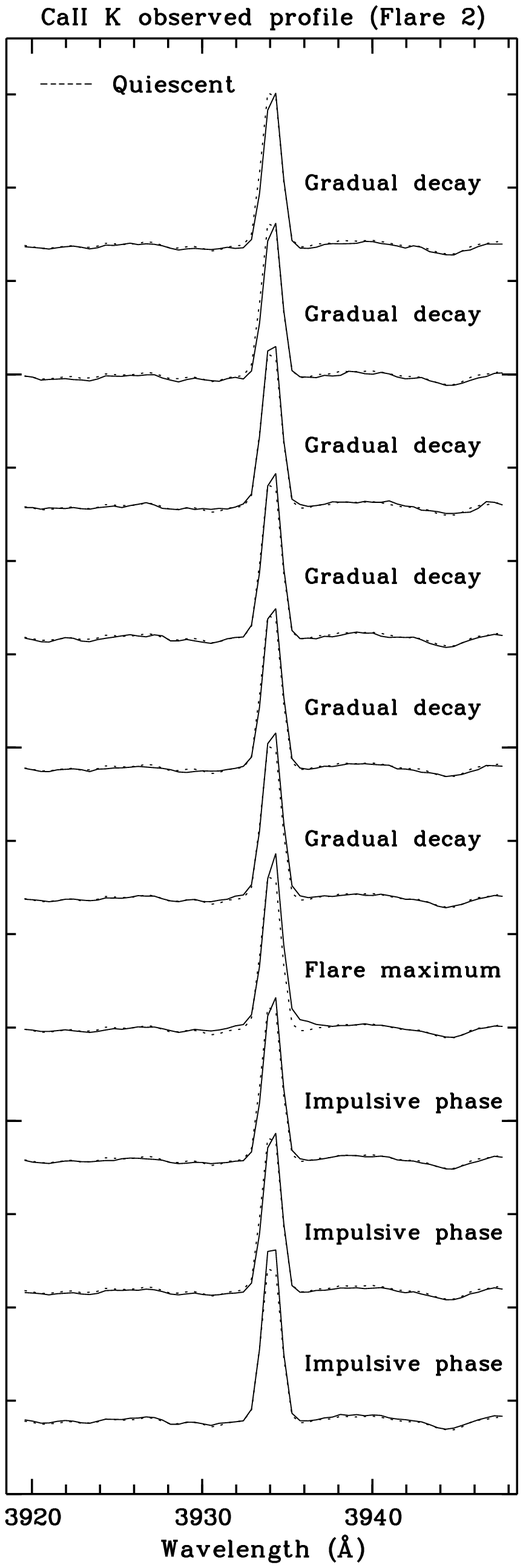,height=14.5cm,width=4.1073cm,bbllx=52pt,bblly=28pt,bburx=303pt,bbury=773pt,clip=}}
\caption[ ]{Evolution of the H$_\beta$ (left) and
\mbox{\ion{Ca}{ii} K} (right)
line profiles (solid line) during the strongest observed flare
(flare~2), compared with the quiescent state (dotted line).
\label{fig:profiles}}
\end{center}
\end{figure}

Fig.~\ref{fig:bisectors} shows, as an example, the 
bisector of the H$\beta$ and \ion{Ca}{ii}~K lines during the \mbox{flare 2}.
The bisector of a line is defined
as the middle points
of the line profile taking points of equal intensity at both
sides of the line \citep{Toner1988,jls03}.

The Balmer lines are broadened during the early-flare
phases and the broadening decreases along the
flare evolution (see Fig.~\ref{fig:profiles}).
Although
the emission rises in both wings of the Balmer lines,
the observed enhancement
and broadening are bigger
in the red wing. Therefore the Balmer lines show
a red asymmetry during the detected flares, and the largest
asymmetry is observed at flare maximum (see also the
behaviour of the bisectors
given in Fig.~\ref{fig:bisectors}).
Besides, the stronger 
the flare (flares 2 and 6) the
greater 
this asymmetry is.
The red asymmetry is also
stronger in higher members of the Balmer series.
We have used a two-Gaussian
(narrow=N and broad=B) fit to model the Balmer lines at
flare maxima. The B component appears red-shifted with respect
to N. For example, at the maximum of 
\mbox{flare 2}
the distance between both components increases from 0.4~\AA~in
the case of H$\beta$ up to 1.5~\AA~in H$_8$.
\citet{Doyle88} observed a similar effect during a flare
on \object{\mbox{YZ CMi}}, finding that H$\gamma$ and H$\delta$
showed symmetrically broadened profiles while H$_8$ and H$_9$
showed predominantly red-shifted material. They suggested that within
an exposure time as those used in this work (see Table~\ref{tab:obslog}) 
several downflows, corresponding to different flare kernels which 
brighten successively one after another, may be present in the 
stellar atmosphere.
Every downflow would produce a red-shifted 
contribution to the Balmer lines, which would be larger
for higher members of the Balmer series (see $\S$~\ref{sec:EWRQ}).
In addition, a smaller red asymmetry in the Balmer lines
is also observed during the quiescent state of \object{\mbox{AD Leo}}.


\begin{table}[t!]
\begin{center}
\caption[]{Width at the base of the chromospheric lines in the
quiescent state and at the maximum of 
flare 2.
\label{tab:widthatthebase}}
\scriptsize
\begin{tabular}{ccccccccccccccc}
\noalign{\smallskip}
\hline  \hline
\noalign{\smallskip}

Line & \multicolumn{2}{c}{Width at the base of the line} \\

\noalign{\smallskip}  
\cline{2-3}
\noalign{\smallskip}

  & Quiescent state & Maximum of 
flare 2 \\

\noalign{\smallskip}
\cline{1-3}
\noalign{\smallskip}

H$\beta$  & 5$\pm$1 & 13$\pm$1 \\
H$\gamma$ & 4$\pm$1 & 6$\pm$1 \\
H$\delta$ & 4$\pm$1 & 9$\pm$1 \\ 
\ion{He}{i} 4026~\AA & 1$\pm$1 & 1$\pm$1 \\ 
\ion{Ca}{ii} H + H$\epsilon$ & 4$\pm$1 & 6$\pm$1 \\
\ion{Ca}{ii} K & 3$\pm$1 & 5$\pm$1 \\
H$_8$ & 3$\pm$1 & 6$\pm$1 \\
H$_9$ & 3$\pm$1 & 5$\pm$1 \\
H$_{10}$ & 3$\pm$1 & 3$\pm$1 \\
H$_{11}$ & 3$\pm$1 & 5$\pm$1 \\

\noalign{\smallskip}
\cline{1-3}
\noalign{\smallskip} 
\end{tabular}
\end{center}
\end{table}

\begin{figure*}
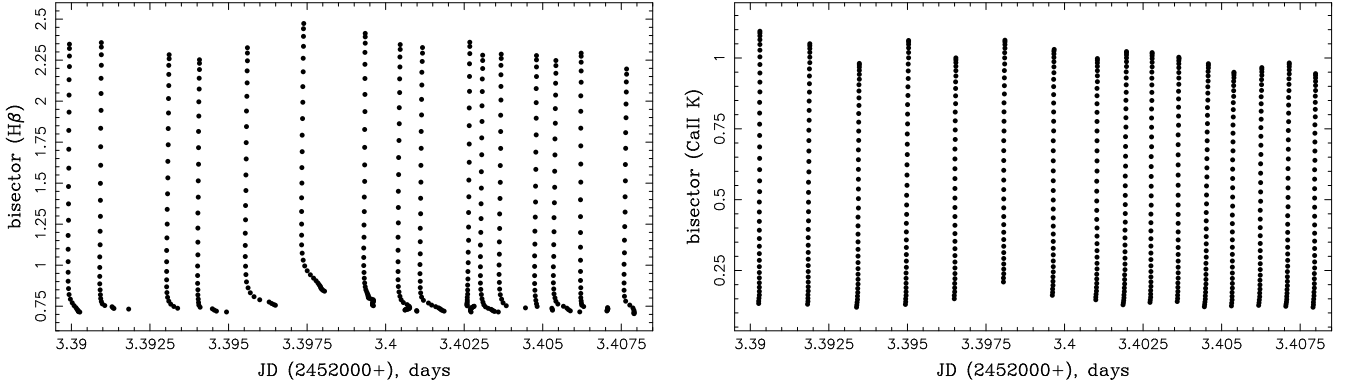

\begin{center}
{\psfig{figure=bisec_Hb_adleo_fulg2_publicacion1.ps,width=8.7cm,bbllx=43pt,bblly=48pt,bburx=550pt,bbury=350pt,clip=}}~~
{\psfig{figure=bisec_CaIIK_adleo_fulg2_publicacion1.ps,width=8.7cm,bbllx=43pt,bblly=48pt,bburx=550pt,bbury=350pt,clip=}}
\caption[ ]{Evolution of the bisector of the H$_\beta$ (left) and \mbox{\ion{Ca}{ii} K} (right)
lines during the strongest flare (flare~2).
Note that only the minimum value of each bisector is placed on the
$JD$ in which its respective spectrum was taken.
\label{fig:bisectors}}
\end{center}
\end{figure*}


The \ion{Ca}{ii}~H \&~K lines
are not usually broadened and
do not show any asymmetry,
excepting the case of the strongest flare, where 
little variations are observed (see Fig.~\ref{fig:profiles} and~\ref{fig:bisectors}).

There are several interpretations for the asymmetries.
Broad Balmer emission wings have been observed during flares
produced by very different kinds of stars,
showing blue or red asymmetries
\citep{Doyle88,Phillips88,Eason92,Houdebine93,Gunn94,Abdul95,Abranin98,Berdyugina98,Montes99,Montes&Ramsey99,jls03}
or no obvious line asymmetries \citep{Hawley&Pettersen91}.
Broadened
profiles and red asymmetries are important constraints on
flare models \citep{Pallavicini1990}.
They can be
attributed to plasma turbulence or mass motions in the flare region
\citep[see][and references therein]{Montes99,Fuhrmeister05}.
In solar flares, most frequently, a red asymmetry is
observed in chromospheric lines.
Taking into account that the flares reported in this work
are like the most typical solar \mbox{flares -- non} \mbox{white-light}
\mbox{flares --} it 
was expected to find red asymmetries like those observed.
However, evidences of blue asymmetries have been also reported
during solar flares \citep{Heinzel94}.
Red asymmetries have been often interpreted as the
result of chromospheric downward condensations (CDC)
\citep[see][and references therein]{Canfield1990}.
In fact, recent line profile calculations
\citep{Gan93,Heinzel94,Ding&Fang97}
show that a CDC can explain both blue and
red asymmetries. On the other hand, evidences of mass
motions have been also reported during stellar flares.
In particular, a large enhancement in the far blue
wings of Balmer lines during the impulsive phase of a stellar
flare was interpreted as high velocity mass ejection
\citep{Houdebine1990} or high velocity chromospheric
evaporation \citep{Gunn1994}, whereas red asymmetries in
Balmer lines were reported by \citet{Houdebine93}
as evidence of CDC. 

Finally, the small red asymmetry observed in the Balmer series during
the quiescent state 
can be therefore interpreted as multiple CDC happening in the stellar 
atmosphere. These CDC may be produced by unresolved
continuous low energy flaring.

\section{Balmer decrement line modeling}
\label{sec:flaresmodelling}

Balmer decrements (flux ratio of higher members to H$\gamma$) have been frequently
used to derive plasma densities and temperatures in the chromosphere of flare stars
\citep{Kunkel70,Gershberg74,Katsova90}. \citet{Jevremovic98} developed a
procedure (BDFP hereafter) to fit the Balmer decrements in order to determine some
physical parameters in the flaring plasma. \citet{Garcia-Alvarez02,Garcia-Alvarez03}
used the BDFP to obtain a detailed trace of physical parameters during several flares.

\begin{table*}[tp!]
\begin{center}
\caption{\small{Physical parameters obtained for the observed flares.
The area and stellar surface percentage covered by these flares are also shown.}
\label{tab:physicalparameters}}
\begin{tabular}{@{}clccccccc@{}}
\noalign{\smallskip}
\hline  \hline
\noalign{\smallskip}
Flare&Date&UT&log $\tau_{Ly \alpha}$&log $n_{\rm e}$ &log $T_{\rm e}$ & log $T_{\rm us}$ &  Area & Surface\\
    &&  &                    &            &            &                          &($\times$ 10$^{19}$~cm$^{2}$) & (\%)\\
\noalign{\smallskip}
\hline
\noalign{\smallskip}
1 &03/04/01 & 00:30 & 3.71&     13.75&    4.31&    4.13& 1.17 & 0.40 \\
2 &03/04/01 & 21:32 & 4.62&     13.79&    4.07&    3.99& 6.79 & 2.30 \\
3 &04/04/01 & 00:55 & 4.18&     14.22&    4.09&    3.90& 1.45 & 0.49 \\
4 &04/04/01 & 01:36 & 3.78&     13.95&    4.30&    3.98& 0.89 & 0.30 \\
5 &04/04/01 & 21:11 & 4.54&     13.79&    4.12&    4.02& 0.76 & 0.26 \\
6 &04/04/01 & 22:08 & 4.43&     14.12&    4.07&    4.04& 3.14 & 1.07 \\
7 &04/04/01 & 23:41 & 4.41&     14.39&    4.09&    4.02& 2.13 & 0.72 \\
8 &05/04/01 & 01:18 & 4.15&     13.78&    4.17&    3.98& 2.12 & 0.72 \\
9 &05/04/01 & 02:25 & 3.60&     13.88&    4.38&    4.09& 0.36 & 0.12 \\
\noalign{\smallskip}
\hline
\noalign{\smallskip}
\end{tabular}
\end{center}
\end{table*}

The BDFP is based on the solution of the radiative transfer equation.
It uses the escape probability technique \citep{Drake80,DrakeUlrich80} and a simplified
picture of the flaring plasma as a slab of hydrogen with an underlying thermal
source of radiation which causes photoionization. 
This source represents a deeper layer in the stellar atmosphere, 
which is exposed to additional heating during flares.
For a detailed description of the physical assumptions 
made by the BDFP, see \mbox{$\S$~4.1} of \citet{Garcia-Alvarez02}.
The BDFP minimizes the difference between the observed and
calculated Balmer decrements using a multi-directional search algorithm
\citep{Torczon91,Torczon92}. This allows us to find the best possible solution
for the Balmer decrements in a four dimensional parameter space, where the
parameters are: electron temperature ($T_{\rm e}$), electron density
($n_{\rm e}$), optical depth in the Ly$\alpha$ line ($\tau_{Ly \alpha}$),
and temperature of the underlying source or background temperature
($T_{\rm us}$). 
Although the temperature of the underlying source could be
a free parameter in our code, we have fixed a lower limit at \mbox{2500~K}
for numerical stability reasons.
The best solution for the Balmer decrements allows us to calculate
the effective thickness of the slab of hydrogen plasma and the total emission measure
per unit volume \citep[see expressions given by][]{DrakeUlrich80}.
With these quantities we are also able to determine the surface area of the 
emitting plasma \mbox{-- volume/thickness --} taking into account that the volume is the ratio of the 
observed flux in the H$\beta$ line to the total emission measure (also in this line) per unit volume.

\subsection{Physical parameters of the observed flares}

We apply the BDFP to the flares observed on \mbox{\object{AD Leo}}.
The Balmer decrements for H$\gamma$, H$\delta$, H$_8$, H$_9$ and H$_{10}$
have been calculated at each flare maximum using the flare-only fluxes 
given in Table~\ref{tab:maximumfluxes}.
We do not include 
higher Balmer lines due to both the difficulty in assigning their
local continuum level and the very low $SNR$ in their spectral region.

Fig.~\ref{fig:BDADLeoMUSICOS01} shows the observed and fitted Balmer decrements
at the maximum of the detected flares.
The strength of an optical flare is related to the slope
of the fit solution for the Balmer decrements
\citep{Garcia-AlvarezThesis}: 
shallower slopes means stronger flares. 
The results after applying the BDFP are listed in Table~\ref{tab:physicalparameters}.
The flares 3, 6 and 7 have a relatively high electron density
(\mbox{$n_{\rm e} >$ 1 $\times$ 10$^{14}$~cm$^{-3}$}) compared with that
found for the other flares (\mbox{6 $\times$ 10$^{13}$~cm$^{-3}$ $<$ $n_{\rm e}<$ 9 $\times$ 10$^{13}$~cm$^{-3}$}).
The flare 3 also has the lowest background temperature (\mbox{$T_{\rm us}$~$\sim$~8000 K})
which means that the ionization balance in this flare is even less radiatively
dominated than in the other ones (\mbox{9500~K~$< T_{\rm us}$ $<$~13500~K}).
The electron temperature of the observed flares ranges from \mbox{12000 K to 24000 K}.
Taking into account that H$_{10}$ seems to be out of the trend
followed by the other Balmer decrements on 
flares 2 and 3 
(see Fig.~\ref{fig:BDADLeoMUSICOS01}) we have
also run the BDFP rejecting this point, obtaining no significant
variations on the results (the logarithmic values
of the physical parameters given in 
Table~\ref{tab:physicalparameters} change a \mbox{mean of 5~\%}).

\begin{figure*}[tp!]
\begin{center}
{\psfig{figure=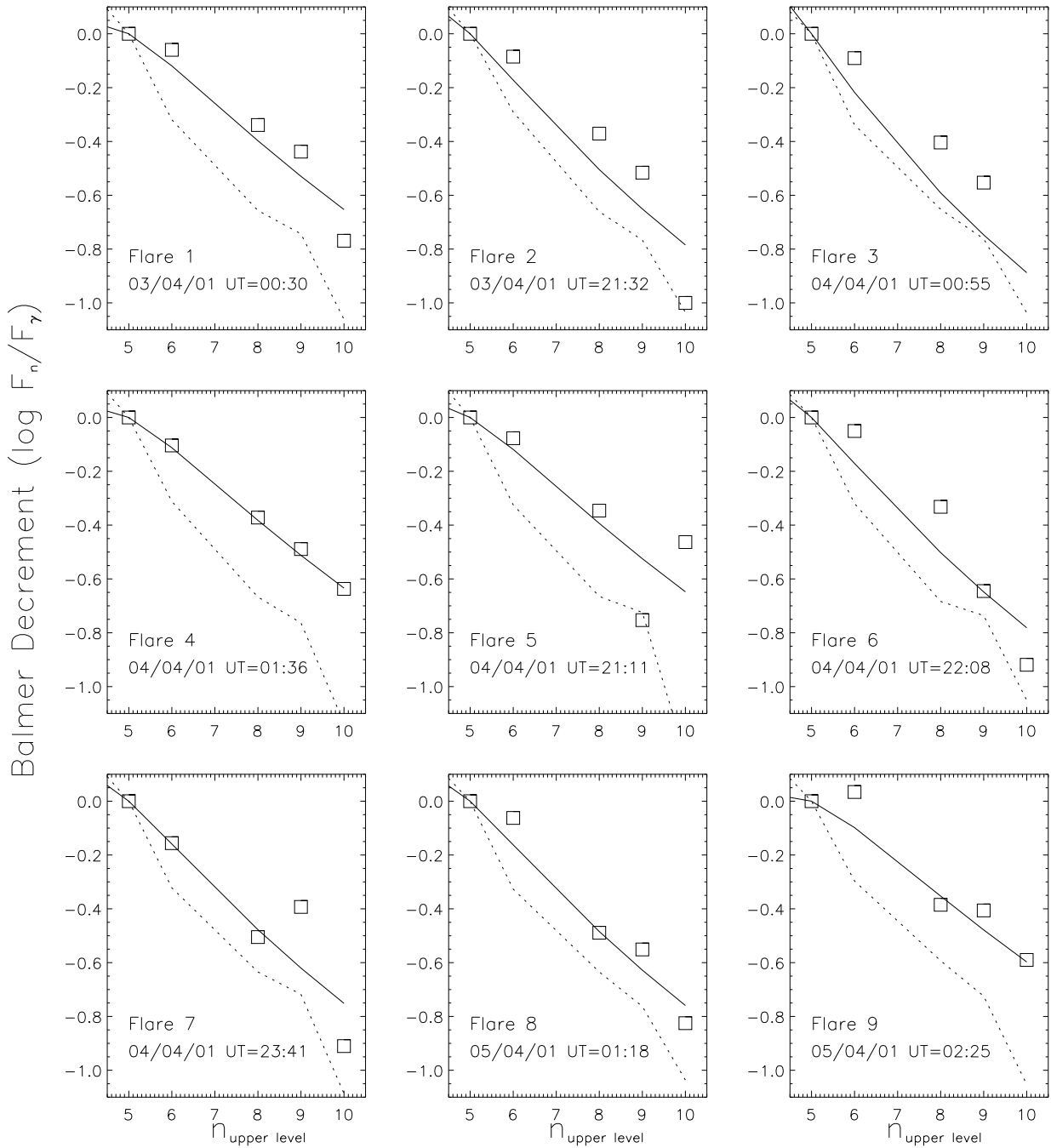,width=16cm}}
\caption[ ]{\small{Observed Balmer decrements (squares) and optimum computed
fits (solid line) at the maximum of the observed flares. 
The pseudo-quiescent state for each flare, as defined in $\S$~\ref{sec:linefluxes},
is plotted as reference (dotted line). Note that the Balmer decrement for
H$_{10}$ is out of the plotted region in flare 3, but it has been taken
into account for doing the fit.}
\label{fig:BDADLeoMUSICOS01}}
\end{center}
\end{figure*}

In addition, we have used the fit solutions for 
calculating the surface area covered by the flaring plasma. 
The values, which are between \mbox{3.6 $\times$ $10^{18}$~cm$^2$} and \mbox{6.8 $\times$ 10$^{19}$~cm$^2$},
are shown in 
Table~\ref{tab:physicalparameters}. 
We find that all the observed flares cover less than 2.3~\% of the projected 
stellar surface ($\pi R^2$).

All the obtained physical parameters are consistent with
previously derived values for stellar flares. These flares are also similar
in size and strength to those analyzed in previous works based on the same code
\citep{Garcia-AlvarezThesis,Garcia-Alvarez02,Garcia-Alvarez03}.

\section{Discussion and conclusions}
\label{sec:conclusions}

The star \mbox{\object{AD Leo}} has been monitored 
using high temporal resolution
during 4 nights \mbox{(2 -- 5 April 2001)}.
More than 600 intermediate resolution spectra
have been analyzed in the optical wavelength range.
Although large variations have not been observed,
we have found
frequent short (duration between \mbox{14~$\pm$~1~and 31~$\pm$~3~min})
and weak (released energy in H$\beta$ between \mbox{1.5  $\times$ 10$^{29}$~and 
8.2 $\times$ 10$^{29}$~erg}) flares.
Most of these flares are even weaker
than those observed on \mbox{\object{AD Leo}} by \citet{Hawley03}.
All the detected flares have been inferred from the variation in the $EW$ of different chromospheric
lines, which has been measured using a very accurate method. 
Given that the continuum does not change during these events, they
can be classified as non \mbox{white-light} flares, which so far represent
the kind of flares least known on stars.

The observed flare activity \mbox{is $>$~0.71~flares/hour}, which is slightly larger
than the values that other authors obtained for this star by using photometry.
As far as we know, it is the first time
that such a high flare frequency is inferred from
the variation of the chromospheric lines.
In fact, \mbox{\object{AD Leo}}
seems to be continuously flaring 
since we have detected 14 moderate flares and a great number of
smaller events that appear superimposed on them.
Given that the energy distribution
of flares was found to be a power law \citep{Datlowe74,Lin84} of the form
\mbox{${dN}/{dE}=kE^{-\alpha}$ -- where} $dN$ is the number of flares (per unit time)
with a total energy (thermal or radiated) in the interval [$E$,~$E+dE$], and
$\alpha$ is greater \mbox{than 0 --}
we can expect 
very weak flares occurring even more frequently than the observed ones.
Therefore our results can be interpreted as an additional evidence     
of the important role that flares can play as heating agents of the outer
atmospheric stellar layers \citep{Gudel97,Audard00,Kashyap02,Gudel03,Arzner04}.

A total of 14 flares have been studied in detail.
The Balmer lines allow to distinguish
two different morphologies:
some flares show a gradual decay much longer than
the impulsive phase while another ones are less impulsive.
This resembles the two main types of solar flares (eruptive and confined) described by \cite{Pallavicini77}.
However, the \mbox{\ion{Ca}{ii}~H \& K} lines always show the same behaviour and their evolution
is less impulsive than that found for the Balmer lines.
The two maxima and/or weak peaks, observed sometimes during the
detected flares, can be interpreted as the succession of 
different magnetic reconnection processes, which would be originated as
consequence of the disturbance produced by the original flare, as \citet{Kopp76} suggested for the Sun.
The \mbox{\ion{Ca}{ii}~H \& K} lines reach the flare maximum after the Balmer series.
It seems that the time delay is higher when the impulsive phase of the Balmer
lines is shorter. The moment in which a line reaches its maximum is related to the
temperature that characterizes the formation of the line and, therefore,
is also related to the height where the line is formed.
During the detected flares, 
the relative increase observed in the emission of the Balmer 
lines is greater for lower wavelengths.
We have also found that, not only the detected
flares are like the most typical solar flares (non \mbox{white-light} flares),
the asymmetries observed in chromospheric lines (red asymmetries) are also like
the most frequent asymmetries observed in the Sun during flares. 
The detected broad Balmer
emission wings and red asymmetries
can be attributed to plasma turbulence, mass motions or CDC. The \mbox{\ion{Ca}{ii}~H \& K}
lines seem to be less affected by flare events and their broadening
is negligible.
A small
red asymmetry in the Balmer series is also observed during the quiescent state,
which could be interpreted as multiple CDC probably due to continuous flaring
of very low energy.

\begin{figure}[t!]
\begin{center}
{\psfig{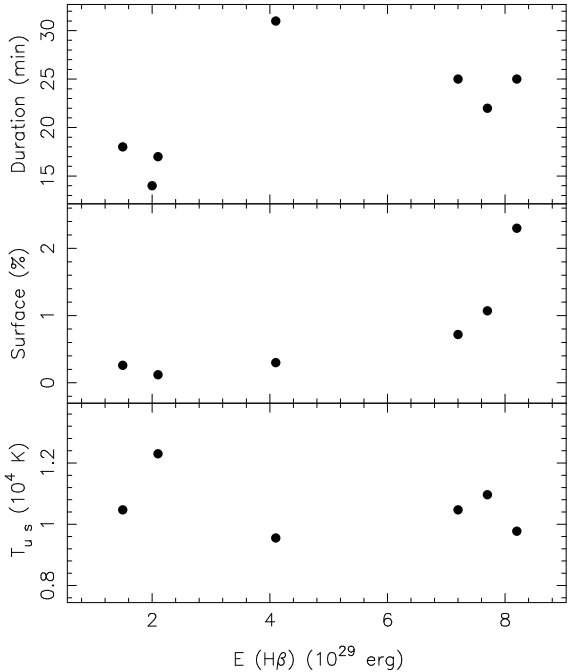}}
\caption[ ]{Flare duration, stellar surface and temperature of the
underlying source at flare maximum vs. energy released
in H$\beta$ during the detected flares.
\label{fig:flarerelationsL}}
\end{center}
\end{figure}

\begin{figure}[t!]
\begin{center}
{\psfig{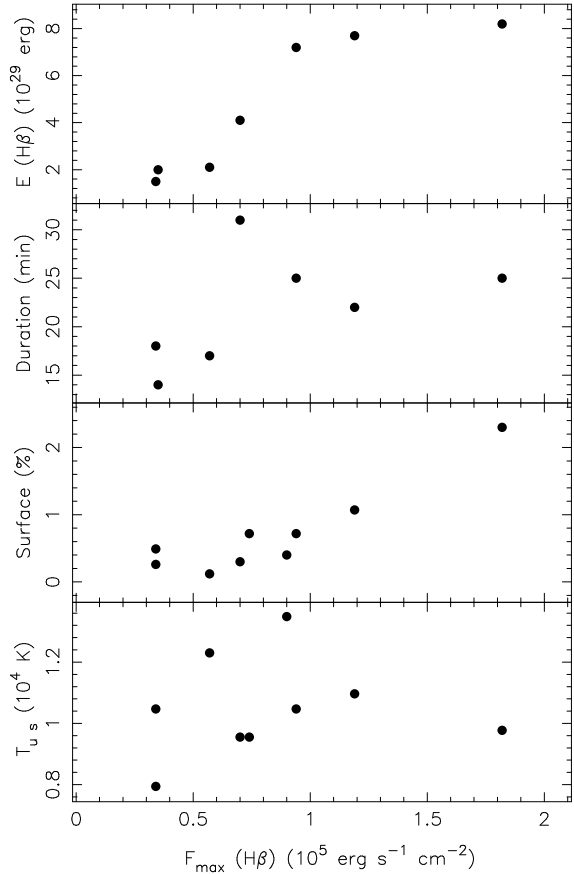}}
\caption[ ]{As Fig.~\ref{fig:flarerelationsL} but vs.
the flux emitted in H$\beta$ at flare maximum.
The top panel has been added to show the relation between this
flux and the energy released in H$\beta$.
\label{fig:flarerelationsF}}
\end{center}
\end{figure}

In addition, we have used the Balmer decrements as a tracer of physical parameters
during flares by using the model developed by \citet{Jevremovic98}. The
physical parameters of the flaring plasma 
(electron density,
electron temperature, optical thickness and temperature of the underlying source),
as well as the covered stellar surface, have been obtained.
The electron densities found for the analyzed flares
(\mbox{6 $\times$ 10$^{13}$~cm$^{-3}$ -- 2 $\times$ 10$^{14}$~cm$^{-3}$})
are in general agreement with those that other authors find by using
semi-empirical chromospheric modeling \citep{1992ApJS...78..565H,1992ApJS...81..885H,1996A&A...310..245M}.
The electron temperatures are between \mbox{12000 K and 24000 K}.
The temperature of the background source ranges from \mbox{8000 K to 13500 K}.
All the obtained physical parameters are consistent with previously
derived values for stellar flares. The \mbox{areas -- no} larger than 2.3~\% of the projected 
stellar \mbox{surface -- are} comparable with the size of other solar and 
stellar flares \citep{Tandberg-Hanssen88,Garcia-AlvarezThesis,Garcia-Alvarez02,Garcia-Alvarez03}.

We have also analyzed the relationships between the flare parameters. 
The released energy is correlated 
with the flare duration and the area covered by the flaring plasma, but not
with the temperature of the underlying source (see Fig.~\ref{fig:flarerelationsL}).
These results are in general agreement with those found by
\citet{Hawley03} using only 4 flares and a different method for
obtaining the physical parameters. We have found a clear relation
between the released energy and the flux at flare maximum 
(see Fig.~\ref{fig:flarerelationsF}). Besides, the higher the flux,
the longer the flare. This flux 
is also correlated with the area covered by the emitting plasma at flare
maximum, but not with the temperature of the underlying source 
(Fig.~\ref{fig:flarerelationsF}). No correlations between the area, 
temperature and duration have been \mbox{found: i.e. these} parameters seem to be independent.
The magnetic geometry seems to be a very important factor in flares: firstly, the
flare duration can be related to the loop length, as in X-rays \citep[see review by][]{2002ASPC..277..103R},
in the sense that a fast decay implies a short loop and a slow decay a large loop; and secondly,
the flare area can be related to the loop width,
because the surface covered by the Balmer emitting plasma would be the 
projected area 
of the flaring loops. On the other hand, the temperature of
the underlying source could be related to the depth of the layer reached by the flare accelerated particles, 
which is also related to the energy released through magnetic reconnection. For explaining the
fact that the temperature of the underlying source seems to be well-defined and independent of the flare energy (see Fig.~\ref{fig:flarerelationsL} and \ref{fig:flarerelationsF}), we suggest that
larger energies may imply more energetic particles and therefore larger depths, but
not necessarily higher temperatures.

This work has extended the current number of stellar flares analyzed
using high temporal resolution and good quality spectroscopic 
observations. However, there is still need for new 
data of this type to trace the behaviour of the physical parameters
throughout these events. This will help us to understand the nature of
flares on dMe stars. 
Finally, higher spectral resolution would be also desirable in order
to study the changes that take place in stellar atmospheres during flares. 

\begin{acknowledgements}
This work has been supported by the Universidad Complutense de Madrid and the
Ministerio de Educaci\'on y Ciencia (Spain), under the grant
AYA2004-03749 (Programa Nacional de Astronom\'{\i}a y Astrof\'{\i}sica).
Research at Armagh Observatory is grant-aided by the Department of
Culture, Arts and Leisure for N. Ireland.
ICC acknowledges support from MEC under AP2001-0475. JLS acknowledges support 
by the Marie Curie Fellowship Contract No MTKD-CT-2004-002769.
We thank the referee for useful comments which have contributed to improve
the manuscript.
\end{acknowledgements}

\bibliographystyle{apj} 
\bibliography{publicacion1}

\end{document}